\documentclass{emulateapj}
\usepackage{gensymb}
\usepackage{upgreek}
\usepackage{color}
\def\spitz{{\it Spitzer }}

\def\mjup{$M_{\rm Jup}$}

\begin{document}

\shortauthors{Esplin et al.}
\shorttitle{Survey for Brown Dwarfs in Chamaeleon I}

\title{A survey for planetary-mass brown dwarfs in the Chamaeleon I star-forming region\altaffilmark{1}}

\author{
T. L. Esplin\altaffilmark{2}, 
K. L. Luhman\altaffilmark{2,3}, 
J. K. Faherty\altaffilmark{4},
E. E. Mamajek\altaffilmark{5,6}, 
and J. J. Bochanski\altaffilmark{7}
}

\altaffiltext{1}{
Based on observations made with the {\it Spitzer Space Telescope}, the NASA/ESA 
{\it Hubble Space Telescope}, Gemini Observatory, the ESO Telescopes at 
Paranal Observatory, Magellan Observatory, 
the Cerro Tololo Inter-American Observatory, and the ESA {\it Gaia} mission.}
\altaffiltext{2}{Department of Astronomy and Astrophysics, The Pennsylvania
State University, University Park, PA 16802, USA; taran.esplin@psu.edu}
\altaffiltext{3}{Center for Exoplanets and Habitable Worlds,
The Pennsylvania State University, University Park, PA 16802, USA}
\altaffiltext{4}{American Museum of Natural History, New York, NY 10024, USA}
\altaffiltext{5}{Jet Propulsion Laboratory, California Institute of Technology, M/S 321-100, 4800 Oak Grove Drive, Pasadena, CA 91109, USA}
\altaffiltext{6}{Department of Physics and Astronomy, University of Rochester, Rochester, NY 14627-0171, USA}
\altaffiltext{7}{Rider University, 2083 Lawrenceville Rd., Lawrenceville, NJ 
08648, USA}

\begin{abstract}
We have performed a search for planetary-mass brown dwarfs in the Chamaeleon~I
star-forming region using proper motions and photometry measured from optical
and infrared images from the {\it Spitzer Space Telescope}, the
{\it Hubble Space Telescope}, and ground-based facilities. 
Through near-infrared spectroscopy at Gemini Observatory,
we have confirmed six of the candidates as new late-type members of 
Chamaeleon~I ($\geq$M8). One of these objects, Cha J11110675$-$7636030, 
has the faintest extinction-corrected $M_K$ among known members, which 
corresponds to a mass of 3--6 M$_{\rm Jup}$ according to evolutionary models.
That object and two other new members have redder mid-IR colors than
young photospheres at $\leq$M9.5, which may indicate the presence of disks. 
However, since those objects may be later than M9.5 and the mid-IR colors of
young photospheres are ill-defined at those types, we cannot determine
conclusively whether color excesses from disks are present. 
If Cha J11110675$-$7636030 does have a disk, it would be a contender for the
least-massive known brown dwarf with a disk. Since the new brown
dwarfs that we have found extend below our completeness limit of
6--10~M$_{\rm Jup}$, deeper observations are needed to measure the minimum
mass of the initial mass function in Chamaeleon~I.

\end{abstract}

\keywords{
planetary systems: protoplanetary disks ---
stars: formation ---
stars: low-mass, brown dwarfs ---
stars: luminosity function, mass function --
stars: pre-main sequence}

\section{Introduction}

The distribution of mass in a population of young stars and brown dwarfs 
(i.e., the initial mass function; IMF) is determined by the process of star
formation. As a result, measurements of the IMF can be used to test star
formation theories. For example, these theories predict 
minimum masses of the IMF that span a wide range of values 
\citep[1--100 \mjup,][references therein]{low76,lar92,whi07}. 
Constraints on the IMF's minimum mass have been provided by surveys for
brown dwarfs in the solar neighborhood \citep[][]{kir12},
young moving groups \citep[][]{gag15}, open clusters \citep[][]{mor04}, 
and star-forming regions \citep[][]{luh09,pen12,lod13,alv13}. 
The most thorough and sensitive surveys have had completeness limits of 
10--20 \mjup\ and have detected objects with masses as low as $\sim$5 \mjup, 
which indicates that the minimum mass of the IMF has not been detected. 

To significantly improve constraints on the IMF's minimum mass,
we need to be able to detect brown dwarfs well below 10~\mjup\ in 
relatively large stellar populations.
This is most easily done in the nearest and richest star-forming regions. 
For instance, according to theoretical evolutionary models \citep{bur97,bur03},
a 2~\mjup\ brown dwarf at an age of 1~Gyr can be detected within only
$\sim$3~pc with the most sensitive wide-field survey for older (colder)
brown dwarfs \citep{wri10}, 
whereas the same object at 1~Myr can be detected with ground based telescopes 
out to the distances of nearby star-forming regions (150--300~pc).
The Chamaeleon I (Cha~I) star-forming region is one of the most promising sites 
for measuring the minimum mass of the IMF because it is rich (N $\sim240$
members), moderately compact (D~$\sim$~$30\arcmin$), and nearby 
\citep[160--170 pc,][]{luhm08}. The current census of the cluster contains
42 members with spectral types later than M6
\citep[$<$0.1~M$_\odot$,][]{neu99,com99,com00,com04,luhm04,luhm07,luh04,luh06,luh08,luhM08,sch08},
where the faintest of which should have a mass of 4--10 \mjup\ \citep{luh08}. 
\cite{muz11,muz15} also obtained spectra of a large number of candidate 
brown dwarfs, but none of them were found to be new members. 

To continue searching for the minimum mass of the IMF in Cha~I,
we have conducted a deep survey for brown dwarfs in a 
wider region than considered in previous studies. 
In our presentation of this survey, we begin
by compiling all known members of Cha~I (Section 2). 
Next, we measure proper motions of stars projected against Cha~I using 
multiple epochs of images from the {\it Hubble Space Telescope} ({\it HST})
and the {\it Spitzer Space Telescope} \citep[][Section 3]{wer04} and we
construct color-magnitude diagrams from optical and near-infrared (IR) images
of the cluster (Section 4). We identify candidate members based on those 
proper motions and color-magnitude diagrams (Section 5) and we describe 
near-IR spectroscopy of the best candidates,
which is used to confirm their youth and cool nature (Section 6).
We conclude by deriving an updated IMF for the cluster (Section 7)
and determining if the new members harbor circumstellar disks based on
mid-IR photometry from {\it Spitzer} (Section 8).

\section{Known Members of Chamaeleon~I}
\label{sec:cens}

We have compiled a catalog of the previously known members of Cha~I.
We began with the census from \cite{luhm08}, which contained 237 members.
Objects appeared as individual entries if they were 
either resolved in \spitz images or classified spectroscopically,
which is the approach that we adopt here.
Because its proper motion from both the {\it HST} and \spitz\ datasets is consistent with membership
(see Section \ref{sec:acspm} \& \ref{sec:iracpm}),
we adopt Cha J11100159$-$7738052 as a previously known member,
which was presented in \cite{luhm07}
as a possible field M9 -- L1 dwarf. 
We also include in our census the M9 companion to CT Cha \citep{sch08}
and five new M4 members from \cite{fra15} and \cite{sac17}, 
which consists of 2MASS J10575375$-$7724495\footnote{2MASS J10575375$-$7724495
was independently confirmed in this study, but is treated as a previously
known member.},
2MASS J10563146$-$7618334,
2MASS J11213079$-$7633351,
2MASS J11130450$-$7534369,
and 2MASS J11090915$-$7553477.
\cite{sac17} suggested that 2MASS J10563146$-$7618334 
is a probable member of the $\epsilon$ Cha association based on its proper
motion in the USNO CCD Astrograph Catalog 4 \citep[UCAC4;][]{zac13}. However, 
we find that its proper motion based on the first data release of the
ESA {\it Gaia} mission \citep{gaia16a,gaia16b}
and the Two Micron Point Source Catalog \citep[2MASS;][]{skr06} is
consistent with membership. 
We list the 244 previously known members and the six new members 
from this study in Table \ref{tab:mem}. 

\section{Proper Motions}\label{sec:pm}

\subsection{Hubble/ACS}
\label{sec:acspm}

\cite{luh05} obtained images of a $13\farcm3 \times 16\farcm7$ area of the southern sub-cluster in Cha~I
(see Figure \ref{fig:opfields}) 
with the Wide Field Camera (WFC) aperture of the Advanced Camera for Surveys (ACS) on board {\it HST}.
The observations were performed on 2004 August 21 and 2005 February 16
with the F775W and F850LP filters. 
To measure proper motions for sources detected in those data,
we repeated those observations in the F850LP filter 
on 2009 August 20 and 2011 February 13 through program 11695.

\begin{figure}[h]
\centering
\includegraphics[trim = 0mm 0mm 0mm 0mm, clip=true, scale=.7]{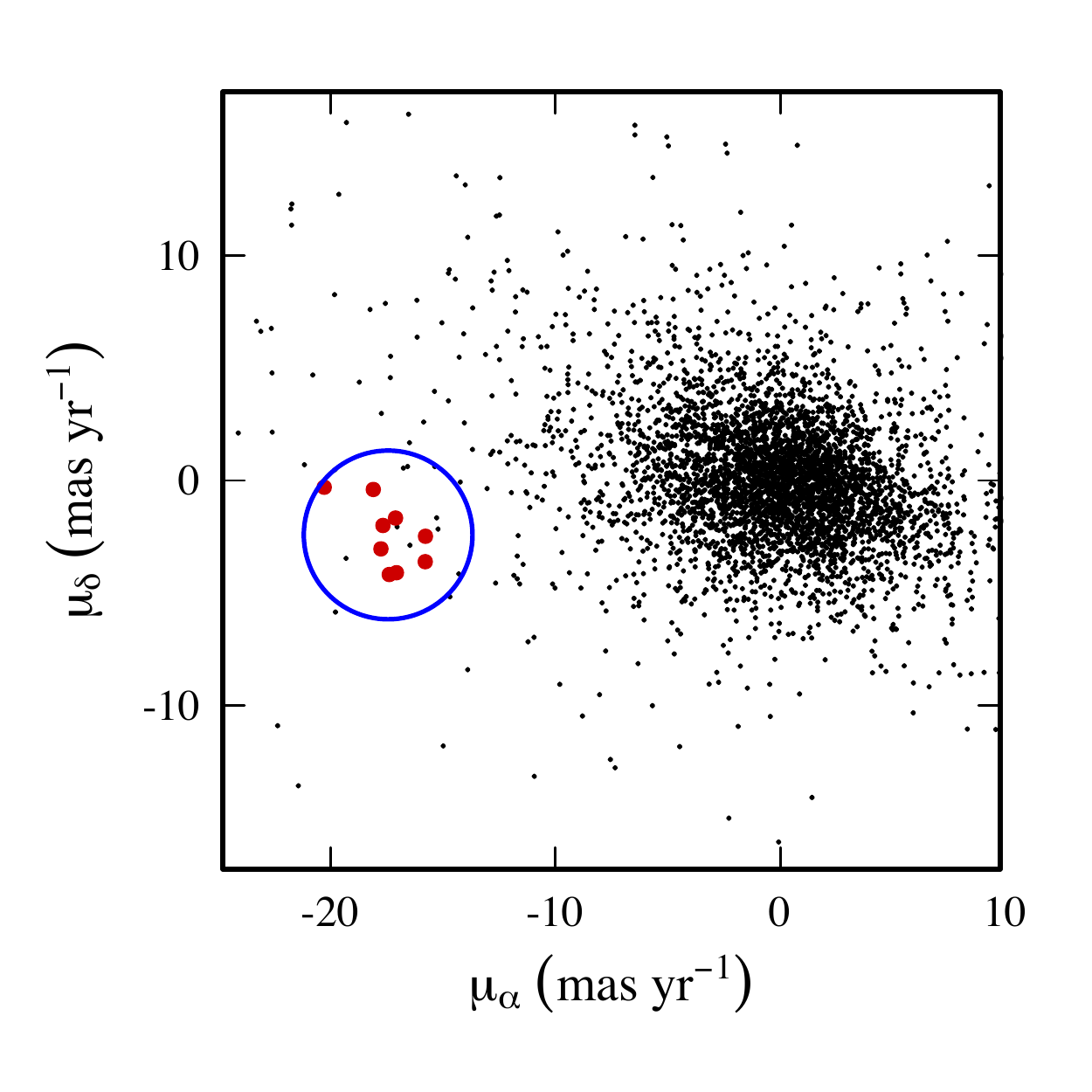}
\caption{
Relative proper motions of known members of Cha~I (large red points) and other 
sources detected in the multi-epoch ACS images (small black points).
We selected sources within 3.75 mas/yr of the mean motion of the known members
(blue circle) as candidate members.
}
\label{fig:acsmu}
\end{figure}

The ACS images were processed and calibrated by the pipeline 
at the Space Telescope Science Institute.
We identified all sources that are not saturated
and measured their pixel coordinates using the IRAF routine {\it starfind}.
We aligned the world coordinate systems of the images at the two epochs 
using sources detected in both sets of data.
We then computed proper motions for sources appearing in both epochs,
which are shown in Figure \ref{fig:acsmu}.
Nine previously known members are detected by ACS and are not saturated. 
Their motions are well-clustered around a mean value of 
$(\mu_\alpha, \mu_\delta) = (-17.4 \pm 1.3, -2.4 \pm 1.5)$ mas/yr.
The known members are well-separated from the vast majority of field stars and galaxies,
which appear near the origin of Figure \ref{fig:acsmu}.
We classify sources within a radius of 3.75~mas/yr of the mean motion of the known members
as candidate members. 

\subsection{Spitzer/IRAC}
\label{sec:iracpm}

Portions of Cha~I have been imaged at several epochs with 
the Infrared Array Camera \citep[IRAC;][]{faz04} on {\it Spitzer}.
Those observations occurred during both the cryogenic and post-cryogenic phases of the mission. 
The cryogenic phase began at launch in August 2003 and continued until 
May 2009 when the liquid helium was depleted. During that time, 
IRAC operated with four 256 $\times$ 256 arrays that collected images in 
broad-band filters at 3.6, 4.5, 5.8, and 8.0~$\mu$m, 
which are denoted as [3.6], [4.5], [5.8], and [8.0]. For each array,
the plate scale was $1\farcs2$ pixel$^{-1}$ and the field of view was $5\farcm2 \times 5\farcm2$.
Point sources within the images have a FWHM of $1\farcs6$--$1\farcs9$
for [3.6]--[8.0]. After the depletion of the cryogen, 
IRAC continued to operate with the [3.6] and [4.5] bands in the post-cryogenic phase. 
\cite{esp16} demonstrated that astrometric precisions
of $\sim$$0\farcs02$ and $0\farcs07$ are possible
for sources in [3.6] and [4.5] images with high and low signal-to-noise ratios (S/N), respectively
\citep[see also][]{dup13,low14}, by measuring source positions with the point-response-function 
fitting routine in the Astronomical Point source EXtractor \citep[APEX;][]{mak05} 
and applying a new distortion correction. 
We have applied these procedures to the multi-epoch IRAC observations of Cha~I.

We retrieved from the {\it Spitzer} archive all [3.6] and [4.5] images
for areas in which at least two epochs of data were available that spanned
several years. Due to their lower sensitivity and resolution, 
we excluded the [5.8] and [8.0] data. 
Astronomical Observing Request IDs (AORs), Program IDs (PIDs), and principle 
investigators for the data are listed in Table \ref{tab:epochs}.
In Figure \ref{fig:iraccov}, we show the spatial coverage of the six 
epochs with large mosaics. 
Epochs 2004.3 and 2005.6 are not plotted because they cover only small fields. 

\cite{luh08} and \cite{luhM08} described the observations of the 
epochs from the start of the mission through 2007.4. The observations from
epoch 2013.6 covered a $\sim$2 deg$^2$ field constructed from nine mosaics.
Each individual mosaic consists of a grid of 18 $\times$ 18 frames 
with a step size of $100\farcs0$ between frames.

For each IRAC frame, we measured positions, fluxes ($F_\nu$), and
S/N's for all detected sources with APEX using the default parameters
except for a 5$\times$5 pixel fitting region, as done in \cite{esp16}.
We then corrected those positions for distortion with the corrections
from the R package ``IRACpm'' \citep{rrr,esp16}.
To compensate for the time-dependence of the [3.6] and [4.5] plate scale,
we interpolated between the measured values presented in \cite{esp16}.
Because their astrometry was unreliable,
sources with $F_\nu / ({\rm exposure\ time}) > 0.728$ and $>$0.8216 Jy/s 
were rejected as saturated for [3.6] and [4.5], respectively. 

Because each source appeared at several different positions in an array in the data from 2013.6,
we used self-calibration to estimate relative offsets and orientations between frames for that epoch.
We derived the initial coordinates and orientations for each frame using 
astrometry from the 2MASS Point Source Catalog. 
We then refined these values by iterating the following four steps:
(1) For each source with a median S/N$>20$ from among the various frames,
we calculated the mean position using the current measured values for 
the frame coordinates and orientations. 
(We refer to the sources's mean position as the ``true'' position, and the 
individual detections as the source's ``measured'' positions.)
(2)~With that new astrometric catalog,
we solved for new frame coordinates and orientations using a least-squares algorithm.
(3) We calculated the root mean square (RMS)
of the residual distances between all measured and true positions.
(4) Detections with residual distances $>$ 6 $\times$ RMS were removed as outliers.
We found that three iterations of these steps were sufficient for the 
measured frame coordinates and orientations to converge on a stable solution. 
Those methods were not applied to the data earlier than than 2013.6 because 
the detections of a given source were confined to a small area of an array, making 
the data less suited for self-calibration.
For these epochs, 
we derived the frame coordinates and orientations by matching sources with
median S/N$>20$ to a subset of the final astrometric catalog from epoch 2013.6.
This subset consisted of sources with errors in position
$\leq$ 0$\farcs$1 and flux measurements $\geq$ 200 $\upmu$Jy in either band.
We then iterated the second through fourth steps of the self-calibration
method four times with a stricter outlier rejection (S/N $>2\times$ RMS). 

For each epoch, we measured astrometry for all sources with median S/N $\geq$ 3
by calculating the mean and median position for sources with 3 and $>$3
total detections, respectively.
We adopted the median absolute deviation (MAD) of the measured positions 
as an estimate for positional error. For sources with $>$3 total detections,
we also calculated median positions and errors for the [3.6] and [4.5] bands
individually. If MAD$_{\rm [3.6] or [4.5]} \times 2 <$ MAD$_{\rm total}$ in
either right ascension or declination,
then we adopted that band's position and error in the final catalog.
Sources with $<$3 total detections were omitted.

\begin{figure}[h]
\centering
\includegraphics[trim = 0mm 0mm 0mm 0mm, clip=true, scale=.75]{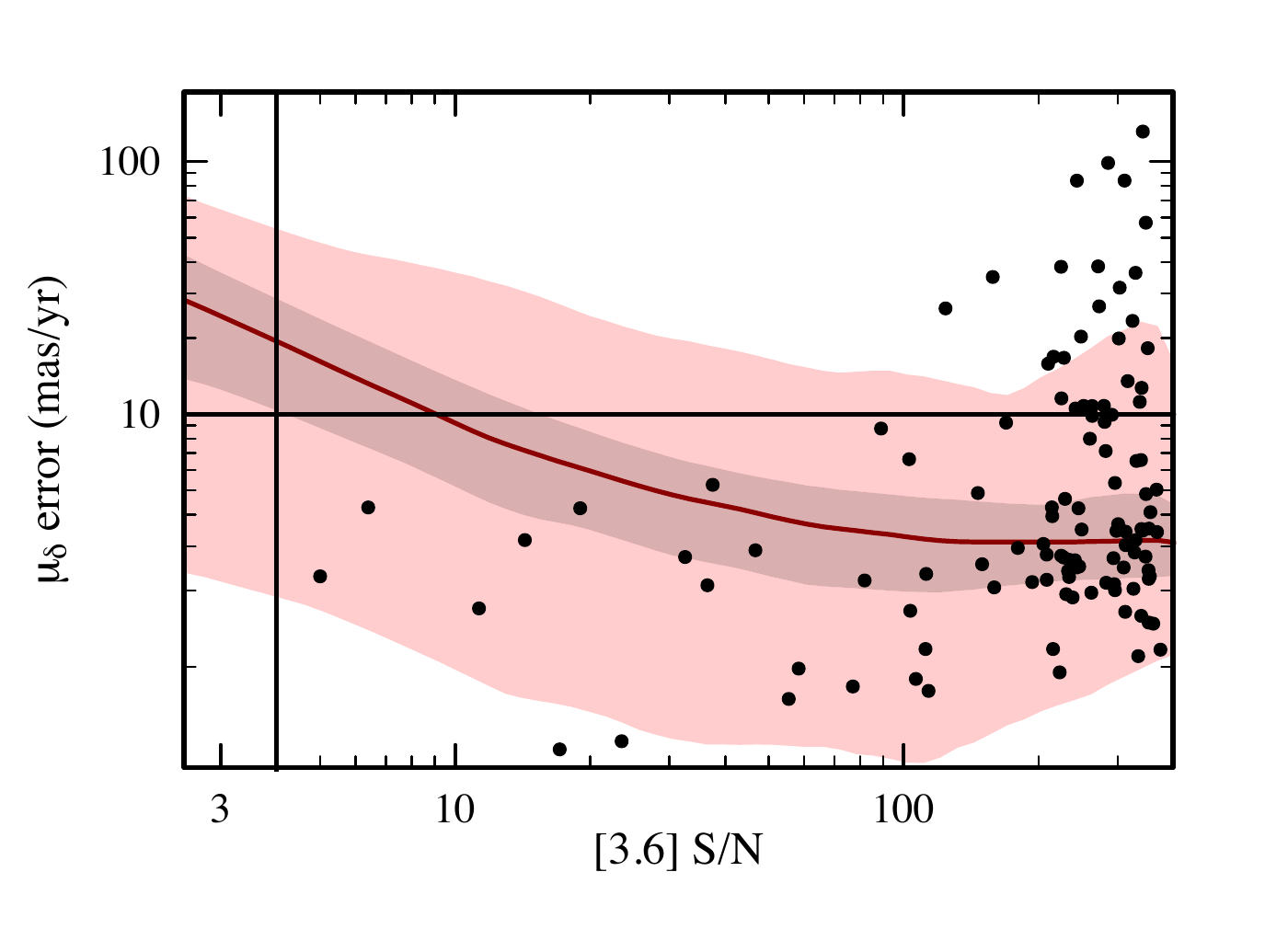}
\caption{
Estimated error of the IRAC proper motion in declination 
as a function of S/N for the known members of Cha~I in the [3.6] band (black points).
The median errors for other sources in the IRAC images are plotted as a solid red line 
with 50\% and 90\% of the errors contained within the darker and lighter shaded regions,
respectively. 
We only consider motions with errors $<$10 mas/yr and S/N $>$ 4 (black lines)
during our candidate selection.
}
\label{fig:iracerror}
\end{figure}

To measure relative proper motions ($\mu$) with the catalogs of all of the 
epochs, we fit each source's astrometry as a function of time with a line.
In Figure \ref{fig:iracerror},
we plot the errors for $\mu_\delta$ as a function of [3.6] S/N for all sources.
We estimated 25\%, 50\%, and 75\% quantiles for these data 
by applying local linear quantile regression with the function 
{\tt lprq} in the R package {\it quantreg} \citep{koe16}. Below $\sim$100 S/N,
the median error increases smoothly with decreasing S/N
and above this limit it is flat at 3 mas/yr.
However, there is a significant increase in the number of 
sources with high error for the highest S/N. 
These star are below the saturation limit and 
do not have anomalous point spread functions (PSFs). 
It appears that APEX, for an unknown reason, 
produces erroneous positions for some bright stars.
To minimize the number of contaminants while maximizing the number of known Cha~I members in our 
proper motion catalog, we only use proper motions with errors in $\mu_\delta \leq$10 mas/yr.
We also omit proper motions for sources with S/N $\leq4$ for [3.6]
so that at least 25\% of sources above a given S/N have motion errors below 10 mas/yr.
For these same reasons, we also omitted proper motion data with
errors in $\mu_\alpha >$ 8~mas/yr and S/N$>6$ for [4.5].

We inspected the IRAC images of the known members of Cha~I 
that exhibited discrepant motions compared to the bulk of the population,
which included 2MASS J11064658$-$7722325, CHXR 15, CHXR 78C, ISO 282,
ISO 86, ISO 143, and 2MASS J11094525$-$7740332.
We found that the PSF of 2MASS J11064658$-$7722325 was extended so we do
not report a proper motion for it.
No anomalies were found in the PSFs of the other six members.
For all of those objects except ISO 86,
proper motions are available from the PPMXL catalog \citep{roe10},
all of which are consistent with membership. 
In addition, each of the six stars shows clear evidence of youth.
Therefore, we retain them as members. 
Given that these objects are among the brighter members,
their discrepant motions may be due to same anomaly mentioned earlier that
produces large errors for some stars at high S/N.

\begin{figure}[h]
\centering
\includegraphics[trim = 0mm 0mm 0mm 0mm, clip=true, scale=.7]{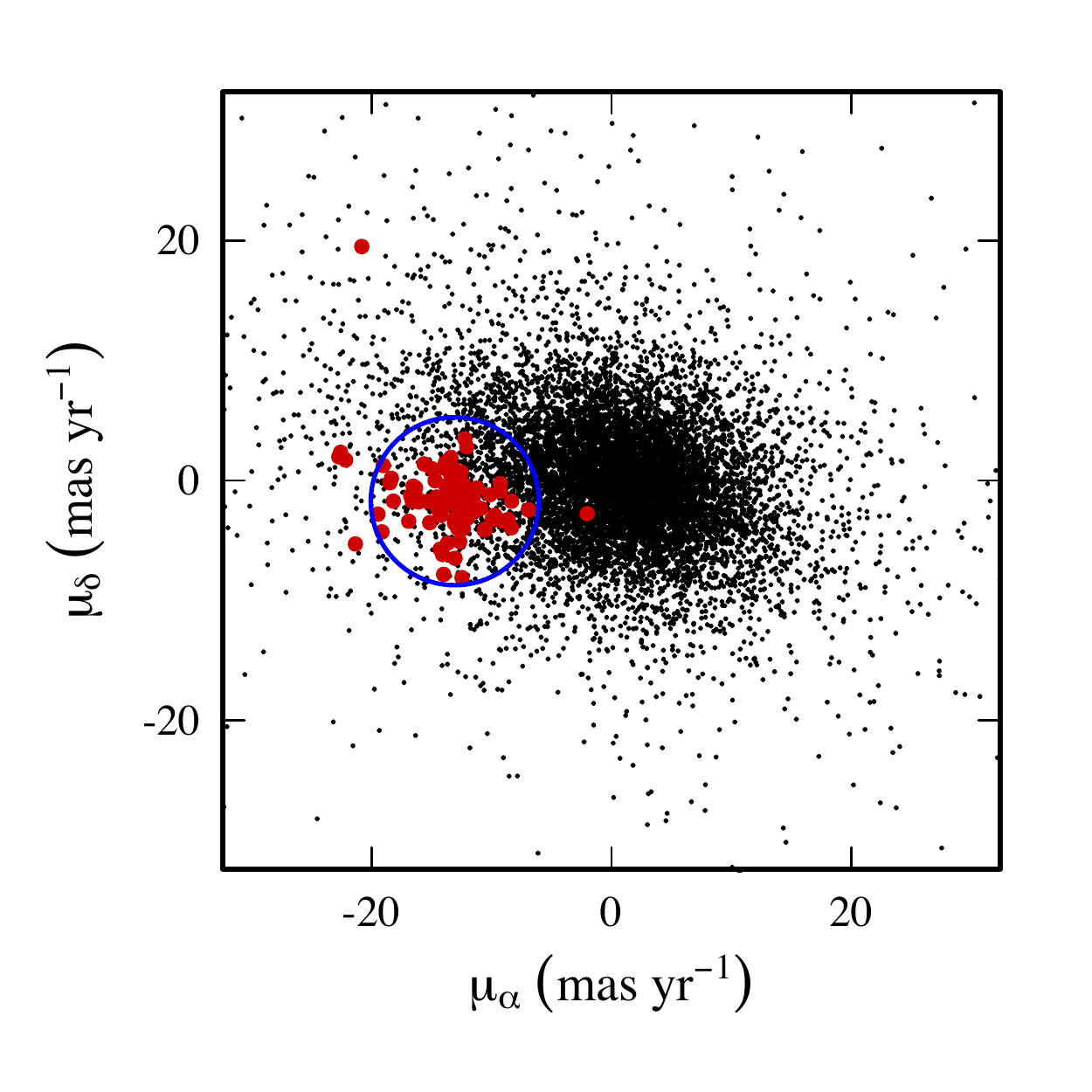}
\caption{
Relative proper motions of known members of Cha~I (large red points) and other 
sources detected in the multi-epoch IRAC images (small black points).
We selected sources within 7.0 mas/yr of the median motion of the known members (blue circle)
as candidate members.
}
\label{fig:iracmu}
\end{figure}

Among the 244 previously known members, 
100 have motions below our adopted limits on proper motion error and S/N,
55 are outside of the region of overlapping IRAC epochs,
38 have errors above our limits,
43 are saturated,
seven are unresolved from bright companions,
and one (Cha--MMS1) was not detected in IRAC.
We include the proper motions for those 100 members with good measurements 
in Table \ref{tab:mem}. Those data have a median value of
$(\mu_\alpha, \mu_\delta) = (-13.0, -1.7)$ mas/yr. The motions for these
members and other sources that satisfy our motion error and S/N criteria 
are plotted in Figure \ref{fig:iracmu}.
The majority of the known members 
are clustered within a radius of 7 mas/yr of the median value. Therefore, 
among the stars whose membership has not been previously determined,
we classify those within that radius as candidate members. 

\section{Photometry}
\label{sec:phot}
\subsection{Previous Data}

Because the field star contamination is high among the sources with IRAC proper motions 
consistent with membership (see Figure \ref{fig:iracmu}),
we used color-magnitude diagrams constructed from
previous and new photometry to identify the most promising candidates. 
These data are also used to identify candidates among sources that lack
proper motion measurements with IRAC. We began by compiling photometry from 
previous surveys for low-mass members of Cha~I and publicly available catalogs,
which consist of F775W and F850LP from ACS/{\it HST}
\citep[see Section \ref{sec:acspm};][]{luh05}; F791W and F850LP from the
Wide Field Planetary Camera 2 (WFPC2) on {\it HST} \citep{tod14};
$I$ from the Inamori Magellan Areal Camera and Spectrograph (IMACS) 
on the Magellan I telescope at Las Campanas Observatory \citep{luhm07};
$Y$, $J$, $H$, and $K_s$ from the Infrared Side Port Imager (ISPI)
at the Cerro Tololo Inter-American Observatory \citep[CTIO;][]{luh05};
$J$, $H$, and $K_s$ from the 2MASS point source catalog;
and $i$ from the Third Release of the 
Deep Near-Infrared Survey of the Southern Sky (DENIS; \citealt{epc99}).
In Figure \ref{fig:opfields}, 
we plot the spatial coverage of the ACS, IMACS, and ISPI images.

\subsection{ISPI}
We obtained images of additional fields of Cha~I with ISPI on the nights of 2008 January 20-23.
This instrument provided a plate scale of 0.3 pixel$^{-1}$ and a field of view (FOV) of
10\farcm25$\times$ 10\farcm25.
These observations consisted of 16 $\times$ 60 s exposures in each of three filters ($J$, $H$, $K_s$)
and in each of four adjacent fields. 
The total area observed corresponds to the middle ISPI field in Figure \ref{fig:opfields}.
The data were reduced in the same manner as the previous ISPI observations \citep{luh05,luhm07}.
The typical FWHM for point sources in these images was 0\farcs95.
These new observations have similar completeness limits as the previous ISPI data, 
which were $J=18.5$, $H=18.25$, and $K_s=17.75$.

\subsection{DECam}

We performed optical imaging of Cha~I with 
the Dark Energy Camera (DECam) at the 4 m Blanco telescope at CTIO.
The instrument contains 62 CCDs with dimensions of 2048 $\times$ 4096 pixels and 
plate scales of 0\farcs27 pixel$^{-1}$. 
The FOV of the entire mosaic has a diameter of $\sim$2\arcdeg.
In an initial set of shallow observations on 2013 May 30, 
we obtained three dithered images in each $i$, $z$, and $Y$
with individual exposures times of 200, 30, and 30~s, respectively. 
We collected deeper data on 2015 March~5
in which the numbers of images and exposure times were 49 $\times$ 150 s ($i$), 
79 $\times$ 100 s ($z$), and 46 $\times$ 200~s ($Y$). 
To extend the dynamic range to brighter magnitudes, 
we also obtained 7 $\times$ 0.7 s images for each filter during that night. 
The spatial coverage the DECam images is shown in Figure \ref{fig:opfields}.

The images were reduced by the DECam pipeline.
The typical FWHM for point sources in these images was 0\farcs8.
We identified all sources in the reduced images and measured their pixel coordinates with {\it starfind}.
We measured aperture photometry for those sources with the task {\it phot} in IRAF 
using a radius of 3.5 pixels and inner and outer radii of 3.5 and 6.5, respectively, for the background 
annulus.
The photometry was calibrated with our previous $I$ and $Y$ data in Cha~I from 
IMACS and ISPI, respectively. 
We calibrated the $z$-band data such that a linear fit
to the sequence of the bluest sources (i.e., $i - [4.5] < 2.5$)
in the color-color diagram $i - [4.5]$ vs. $z - [4.5]$
intersects the origin. 
We estimated the completeness limits of these data based on the magnitudes at which
the logarithm of the number of sources as a function of magnitudes departs from
a linear slope and begins to turn over, which were 22.0, 20.8, and 19.3
for $i$, $z$, and $Y$, respectively.

\subsection{HAWK-I}

\begin{figure}[h]
\centering
\includegraphics[trim = 0mm 0mm 0mm 0mm, clip=true, scale=0.85]{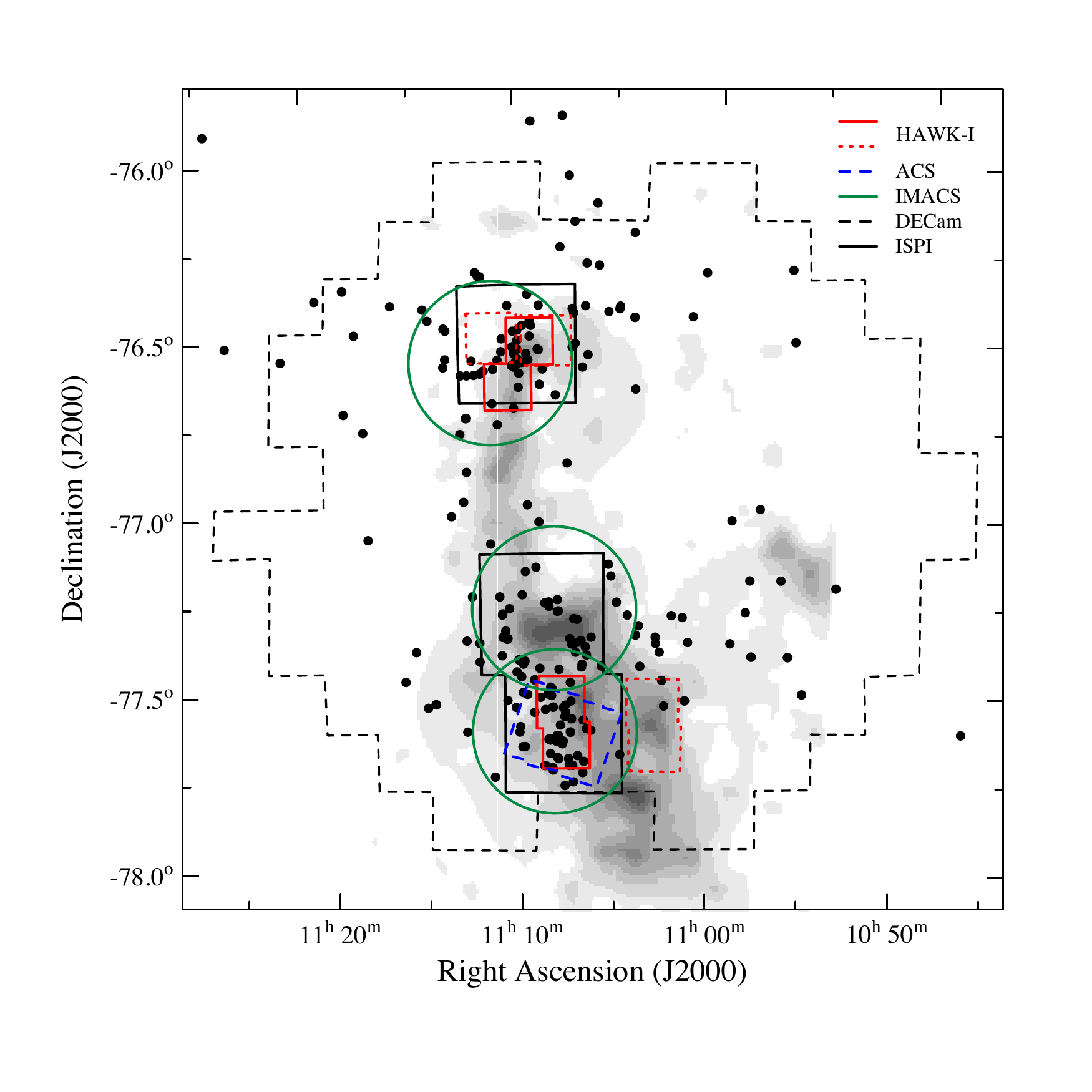}
\caption{
Map of the fields in Cha~I for which we have optical and near-IR photometry
from the following cameras: DECam (black dashed line), 
ACS (blue dashed line), IMACS (green solid line),
shallow HAWK-I (red dotted line), deep HAWK-I (red solid line), 
and ISPI (black solid line).
Data from DENIS and 2MASS are available for the entire region.
The gray scale represents the extinction map from \cite{cam97}.
}
\label{fig:opfields}
\end{figure}

\begin{figure}[h]
\centering
\includegraphics[trim = 0mm 0mm 0mm 0mm, clip=true, scale=0.85]{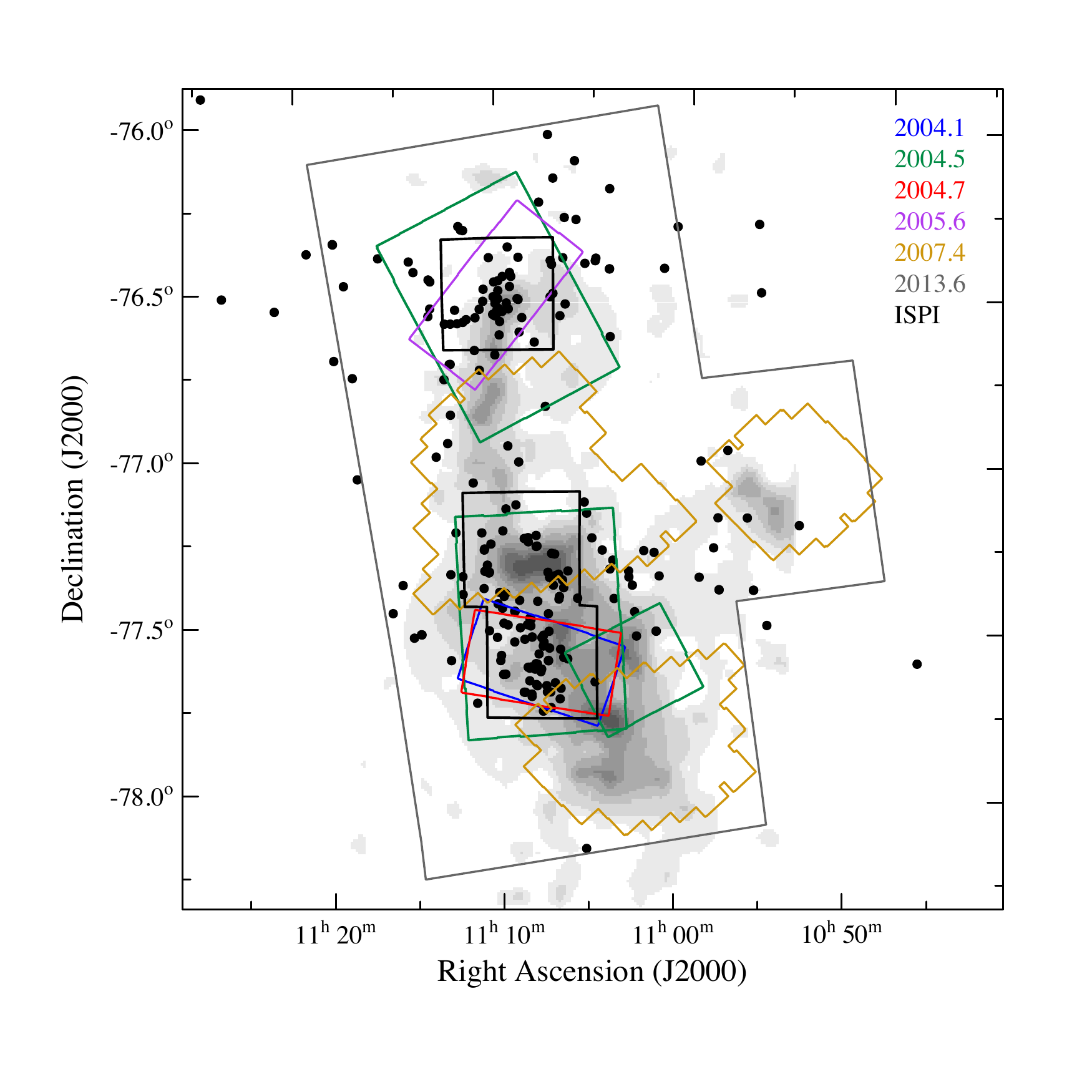}
\caption{
Map of the fields in Cha~I that were imaged by IRAC at multiple epochs
(Table~\ref{tab:epochs}).
The ISPI fields from Figure \ref{fig:opfields} are included for reference.
}
\label{fig:iraccov}
\end{figure}

Two publicly available $K_s$-band imaging programs of Cha~I have been
performed by the High Acuity Wide-field $K$-band Imager (HAWK-I)
on the Unit Telescope 4 of the Very Large Telescope (VLT).
This camera contains four 2048 $\times$ 2048 HAWAII-2RG arrays
which have a plate scale of 0\farcs106 pixel$^{-1}$ 
and a FOV of 7\farcm5 $\times$ 7\farcm5 \citep{kiss08}.
On the nights of 2008 January 24, 27--29, and 31, 
four fields were imaged through program 60.A-9284(L), which are indicated as solid red lines in Figure \ref{fig:opfields}.  In each field, 
25 dithered images were taken each consisting of 15 co-added 2.0 s exposures. 
Four additional fields were imaged 
on the nights of 2010 May 12 and 2010 July 5--6 through program 385.C-0384(A).
Those fields are marked by red dashed lines in Figure \ref{fig:opfields}.
For each field, 200 images were taken with exposure times of 3~s. 

We reduced the individual exposures from HAWK-I by subtracting a dark 
frame and dividing by a flat field image using tasks within IRAF. 
The resulting images were then registered and combined into mosaics.
For most of these images,
the typical FWHM for point sources was 0\farcs8
except for the nights of 2010 July 5--6, when it was 1\farcs5.
For the latter data, 
we measured aperture photometry with a radius of 12 pixels 
and inner and outer radii of 12 and 16, respectively, for the background annulus.
We used values of 6, 6, and 10 for those parameters for the remaining data. 
We measured the WCS for each mosaic using astrometry from 2MASS and our previous images.
The data were flux calibrated with photometry from 2MASS. 
The completeness limits of these data range from $K_s$ = 17.5 to 19.3,
the former corresponding to the data with poor seeing. 

\section{Candidate Selection}\label{sec:selection}

To identify candidate members of Cha~I with the proper motions and photometry described
in the previous two sections, we began by merging the catalogs of sources from IRAC,
ISPI, ACS, 2MASS, DENIS, DECam, HAWK-I, and WFCP2. For each star, when data were available
from multiple camera for a given filter, we adopted the measurement with the smallest error.
We treated the $i$/$I$ filters from DENIS, IMACS, and DECam as a single filter since those data
agreed well with each other. Only data with errors $\leq0.1$~mag were used in our analysis.

\begin{figure}[h]
\centering
\includegraphics[trim = 0mm 0mm 0mm 0mm, clip=true, scale=0.7]{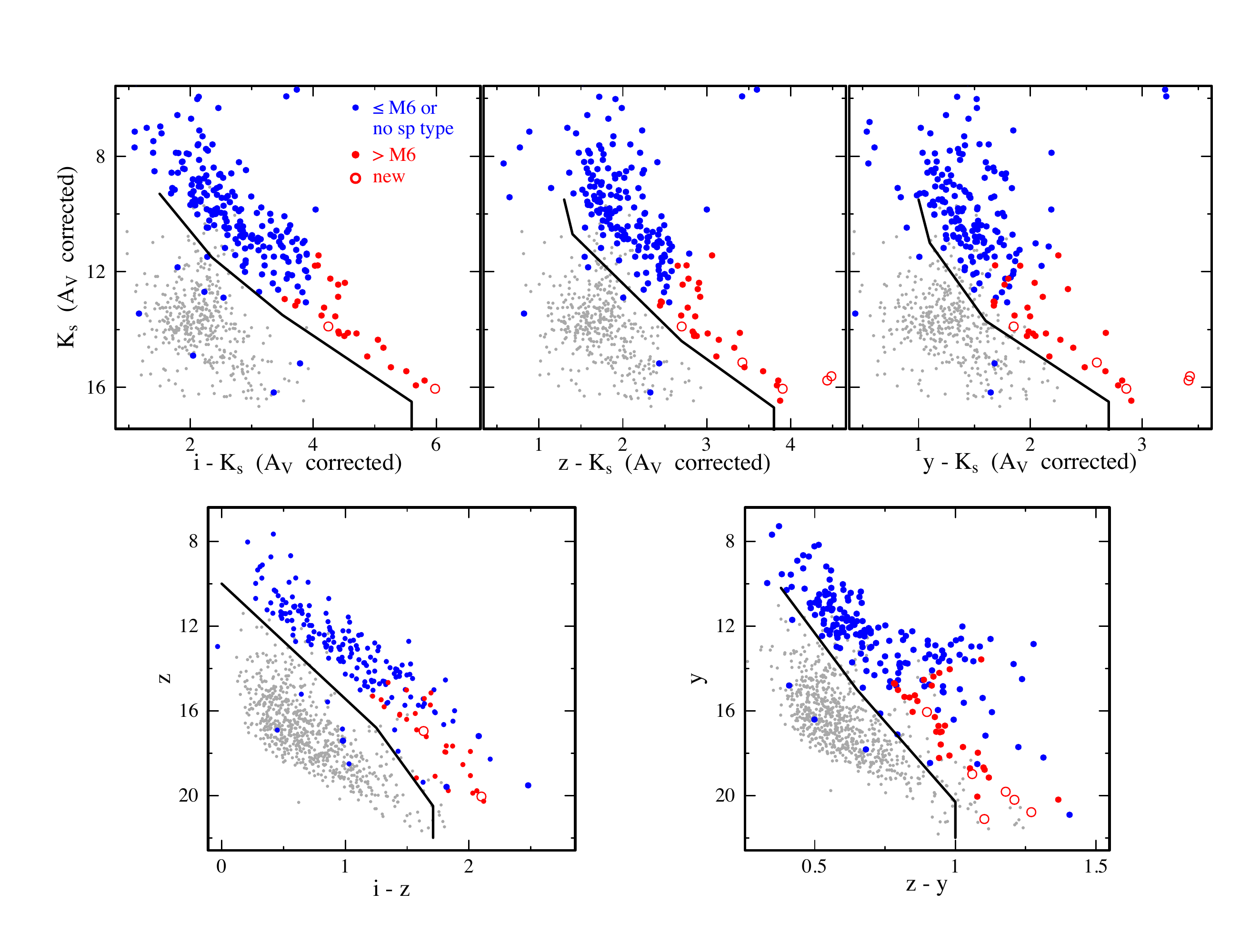}
\caption{
Color-magnitude diagrams for the known members of Cha~I (large filled circles),
new members found in this work (open circles),
and other sources with IRAC motions that are 
consistent with membership (small gray points).
These data are from DECam, ISPI, 2MASS, DENIS, IMACS, and HAWK-I. 
Candidate members have been selected based on positions above the solid boundaries.
}
\label{fig:crit}
\end{figure}

We have used color-magnitude diagrams (CMDs) constructed from our merged catalog 
to select stars that have photometry consistent with membership in Cha~I.
In these diagrams, 
we use $K_s$ as the magnitude because extinction is low in this band and our $K_s$ data offer
good coverage and depth. 
We employ $i - K_s$, $z - K_s$, and $Y - K_s$ because they are sensitive to spectral type.
We can minimize the contamination from reddened background stars 
by estimating and correcting for the extinctions of individual stars in these diagrams \citep{luh03}.
We have estimated the extinction for each star by dereddening its 
positions in $H$ versus $J-H$ to the typical locus 
of young stars at the distance of Cha~I,
which can be approximated by $J-H = 0.68$ for $H < 14.5$ and 
$J-H = 0.128 \times H - 1.176$ for $H \geq 14.5$.
When doing this, we adopt the extinction law from \cite{car89}. 
The resulting extinction-corrected CMDs are shown in Figure \ref{fig:crit}.
To include stars that lack $J$ or $H$, 
we also plot two CMDs that were constructed from optical data alone,
$z$ versus $i - z$ and $Y$ versus $z - Y$.
In each diagram we marked a boundary that follows the lower envelope of the sequence of members. 
A given source is classified as a photometric candidate if it is above a boundary in at least one diagram 
and is not below a boundary in any diagram. 
We also applied the CMDs from \cite{luhm07} and \cite{tod14} to our catalog in the same way.

We selected for spectroscopy objects that were candidates based on either
proper motions (Section \ref{sec:pm}) or CMDs and were not rejected by either
method. We gave higher priority to candidates with faint photometry that was
indicative of lower masses
and that satisfied both our proper motion criteria and multiple CMDs.
We also focused on candidates in the ISPI images because we wished 
to measure an IMF with a well-defined
completeness in those fields (see Section \ref{sec:imf}). 
In addition, we observed one brighter candidate, 2MASS J10575375$-$7724495, 
that exhibits X-ray emission indicative of membership \citep{ing11}. 
The resulting sample contained 11 candidates members of Cha~I.

\section{Spectroscopy of Candidate members}
 \label{sec:spec}

\subsection{Observations}

We performed optical spectroscopy on two targets, 
2MASS J10575375$-$7724495 and 2MASS J11093277$-$7638376,
with the Goodman High Throughput Spectrograph 
at the Southern Astrophysical Research Telescope on 2014 May 7 and 2014 June 16,
respectively. The instrument was operated with the 400~l~mm$^{-1}$ grating
in second order, the GG445 filter, and the 0$\farcs$84 slit, which produced
a wavelength coverage of 5400--9400~\AA\ with resolution of R=1100.
Spectra of six candidates were obtained with FLAMINGOS-2 on the Gemini South
Telescope \citep{eik04} on January 23 and February 2, 3 and 5 in 2015. 
For the brightest target, 2MASS J11124771$-$7737547, the instrument was
configured with the $JH$ grism and filter and the 1$\farcs$08 slit
(0.70--2.04 $\mu$m, R = 300).
For the other five FLAMINGOS-2 targets, 
Cha J11110675$-$7636030, 
Cha J11064106$-$7745040, 
Cha J11105004$-$7721535, 
Cha J11105772$-$7714570, and 
Cha J11104183$-$7633064, 
were observed with the $HK$ grism and filter and the 0$\farcs$72 slit
(1.10--2.65 $\mu$m, R = 450). We observed 2MASS J10543141$-$7710130, 
2MASS J10532815$-$7710268, and 2MASS J11065677$-$7725478 with ARCoIRIS on
the 4 m Blanco telescope at CTIO on the nights of 2016 June 17 and 18
(0.8--2.47 $\mu$m, R=3500).

\begin{figure}[h]
\centering
\includegraphics[trim = 0mm 0mm 0mm 0mm, clip=true, scale=0.8]{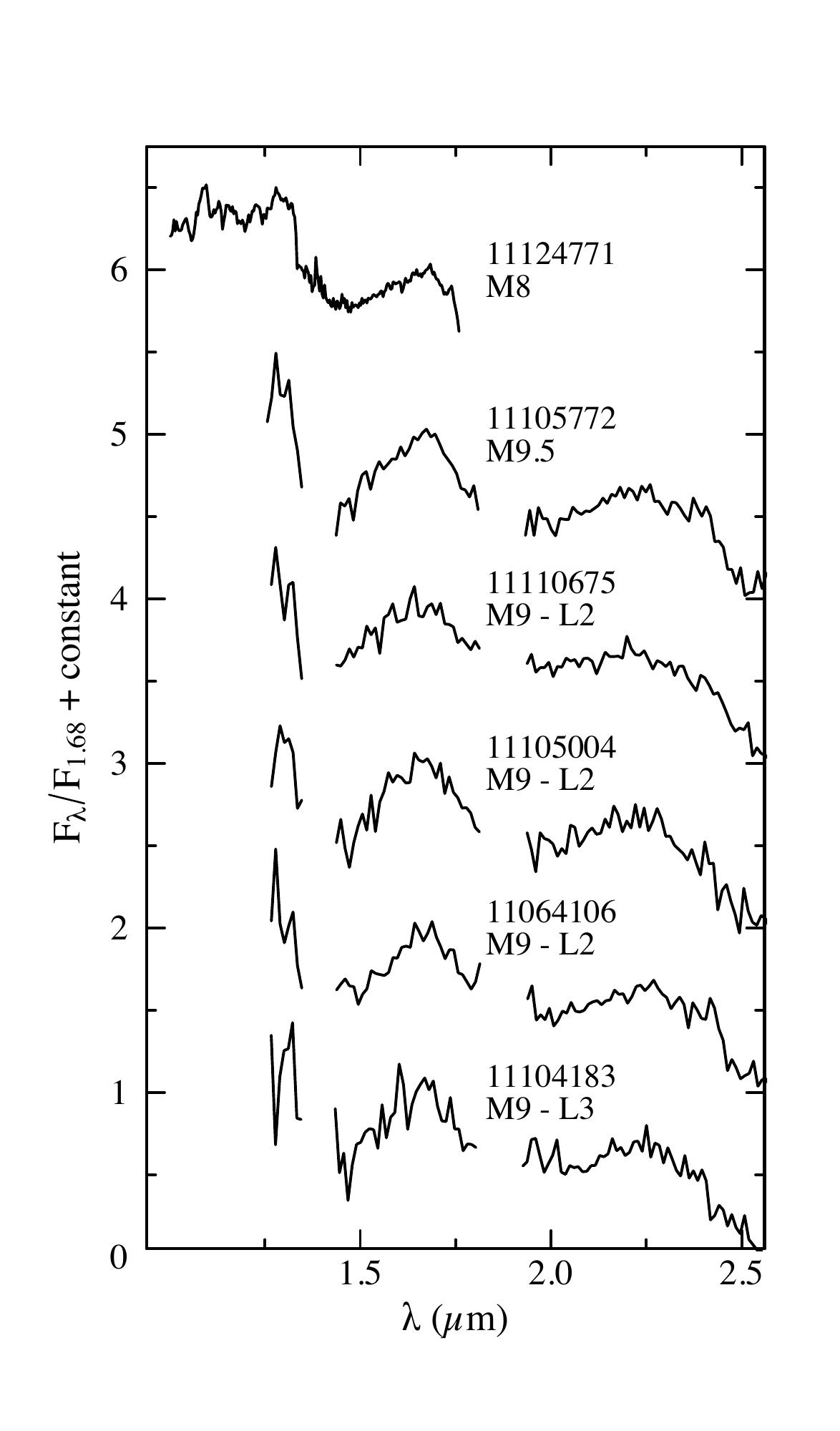}
\caption{
Near-IR spectra of new members of Cha~I. The data have been dereddened
to match standard young brown dwarfs \citep{luh16b}
and binned to a resolution of R = 60 and 30 for the first and remaining
objects, respectively. The data used to create this figure are available.
}
\label{fig:spec}
\end{figure}

For the Goodman and FLAMINGOS-2 spectra, we used routines in IRAF to apply
the flat field correction, extraction of spectra, and wavelength calibration. 
The FLAMINGOS-2 spectra were corrected for telluric absorption using 
a spectrum of an A-type star that was observed at a similar airmass. 
Because the resulting FLAMINGOS-2 spectra had low S/N, we binned the
spectra by a factor of 5 and 15 for data obtained with the $JH$ and $HK$
grisms, respectively. The ARCoIRIS data was reduced with a version
of the Spextool package \citep{cus04} that was modified for use with ARCoIRIS.
The data were corrected for telluric absorption in the manner described
by \cite{vac03}. In Figure \ref{fig:spec},
we show the dereddened spectra of the candidates that we 
classify as new members in the next section. 

\subsection{Spectral Classification}

We have used our spectra to measure the spectral types of our candidates and
to determine whether they show evidence of youth that would indicate membership
in Cha~I. Based on the magnitude of the candidates, they should have spectral
types $\gtrsim$M0 if they are members.
At those types, we distinguished between young objects and field dwarfs with
gravity-sensitive features such as Na~I, K~I, and the shape of the $H$-band
continuum \citep{mar96,luh97,luc01}.
Among the candidates observed with Goodman and ARCoIRIS, 
only 2MASS J10575375$-$7724495 showed evidence of youth.
The other four candidates were background early-type stars or giants. 
All six objects observed with FLAMINGOS-2 were found to be young, late-type
members of Cha~I.
The optical spectrum of 2MASS J10575375$-$7724495 was classified
via comparison to averages of dwarf and giant standards
\citep{luh97,luh98,luh99}.
Most optical spectral types of late-type members of nearby star-forming
regions have been measured with that method.
We arrived at a spectral type of M4 for 2MASS J10575375$-$7724495, which
agrees with the previous classification from \cite{fra15}.
For the new members with near-IR spectra, we measured spectral types
using the standard spectra for young M and L dwarfs from \cite{luh16b}.
The M-type standards were constructed from near-IR data for members of
star-forming regions and young associations ($\lesssim$10~Myr) that have
been classified at optical wavelengths. 
Thus, the near-IR standards from \cite{luh16b} produce spectral types that
are on the same system that has been applied to most known late-type members
of star-forming regions.
The classifications for our new members are presented in Table \ref{tab:spec}.
The uncertainties are $\pm$0.5 subclass unless indicated otherwise. 
Because of degeneracies between spectral type and reddening at late M and L types for young objects \citep{luh16b},
some of the spectral types have large uncertainties.

\begin{figure}[h]
\centering
\includegraphics[trim = 0mm 0mm 0mm 0mm, clip=true, scale=.8]{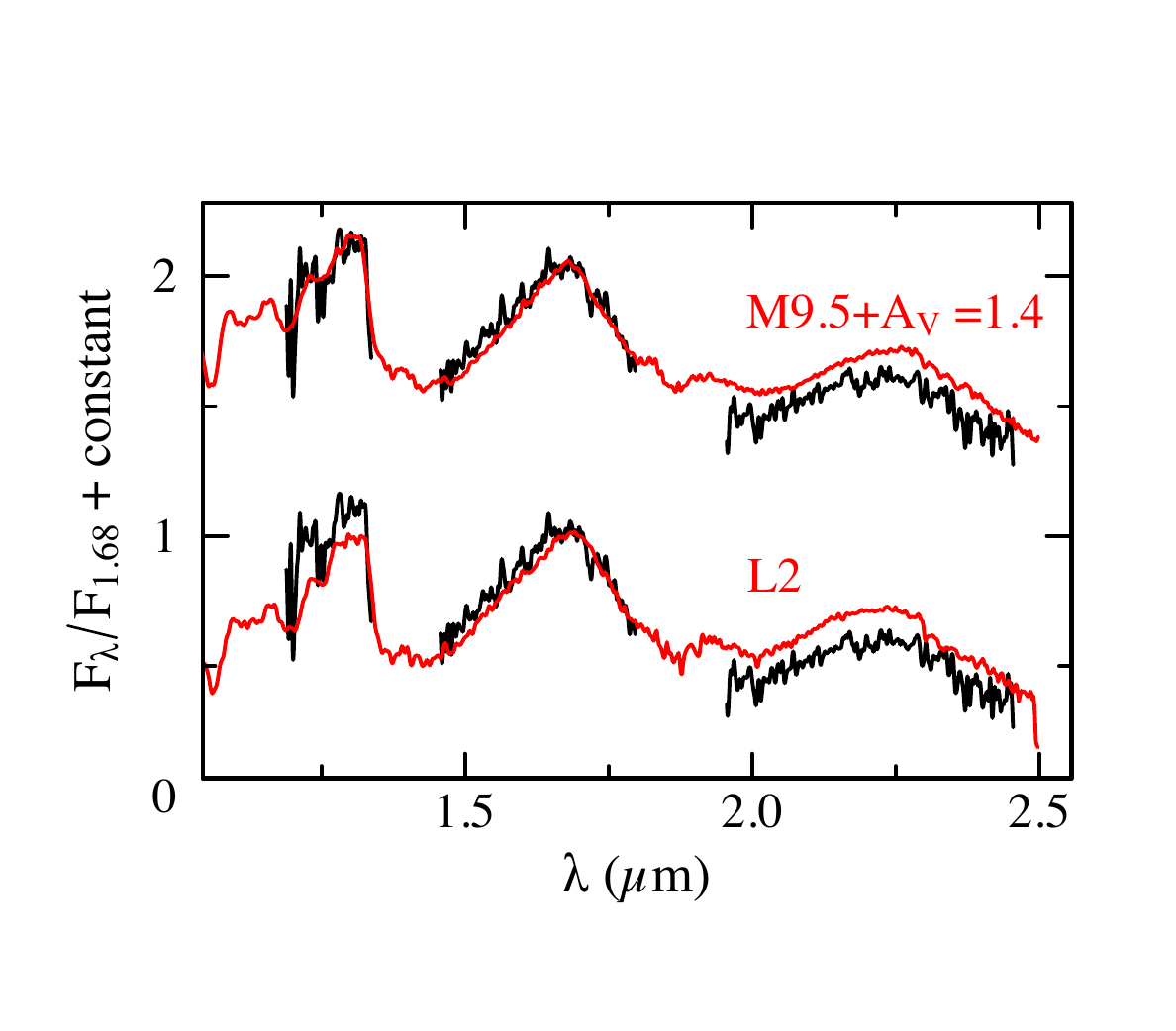}
\caption{
The near-IR spectrum of Cha J11083040$-$7731387 from \cite{luhm07}
(black lines). \cite{muz15} classified this object as L3,
but we find a better match to late M, as illustrated in this comparison
to young M9.5 and L2 standards \citep[red lines,][]{luh16b}. 
A young L3 would be even redder than these standards
and therefore would be much too red to match Cha J11083040$-$7731387.
}
\label{fig:l3}
\end{figure}

In the compilation of the known members of Cha~I in Table~\ref{tab:mem},
we list adopted spectral types based on previous classifications. 
For the previously known members earlier than M9, 
we have adopted the same spectral types as \cite{luhm08}.
We have adopted our classifications of the new members, 
using L0 for the ones with large errors (e.g., M9--L2).
We have reclassified the spectra of members with types later than
M9 \citep{luh06,luhm07} using the standards from \cite{luh16b},
arriving at the following revised classifications:
2MASS J11080609$-$7739406 (M9.5$\pm$0.5), 
Cha J11072647$-$7742408 (M9--L2), 
 Cha J11083040$-$7731387 (M9.5$\pm$0.5),
Cha J11100159$-$7738052 (M9--L3), and 
CHXR 73B (M9.5--L4).
As with the new members, we adopt L0 for those classifications that have
large errors. One of those sources, Cha J11083040$-$7731387, was
classified by \cite{muz15} as L3 based on a low S/N 1.5--2.5~$\mu$m spectrum. 
However, the near-IR spectrum from \citet{luhm07} is bluer than expected
for a young early L dwarf, as shown in Figure \ref{fig:l3}.

\section{Initial Mass Function}

\subsection{Completeness}
\label{sec:comp}

The previous census of members of Cha~I from \cite{luhm07}
had well-defined completeness limits within two fields,
which consisted of a $1\fdg5 \times 0\fdg35$ strip along the center of 
the molecular cloud and the $0\fdg22 \times 0\fdg28$ area imaged by ACS field. 
The limits were $H < 14.5$ for $A_J<1.2$ and $H < 17$ for $A_J<1.4$ 
in those fields, respectively, 
which correspond to masses of $\sim$30~$M_{\rm Jup}$ and $\sim$10~$M_{\rm Jup}$
according to evolutionary models. 
\cite{luhm07} constructed IMFs from the known members within those fields and 
extinction limits, which contained 85 and 34 members. 
To improve the constraints on the minimum mass of the IMF in Cha~I,
we need to obtain a census of members that is complete to faint magnitudes for a field 
that encompasses a large number of members. 
Given these constraints,
we have searched for new members primarily in the ISPI fields 
(0.3~deg$^2$; see Figure \ref{fig:opfields}),
as mentioned in Section \ref{sec:selection}.

\begin{figure}[h]
\centering
\includegraphics[trim = 0mm 10mm 0mm 0mm, clip=true, scale=0.8]{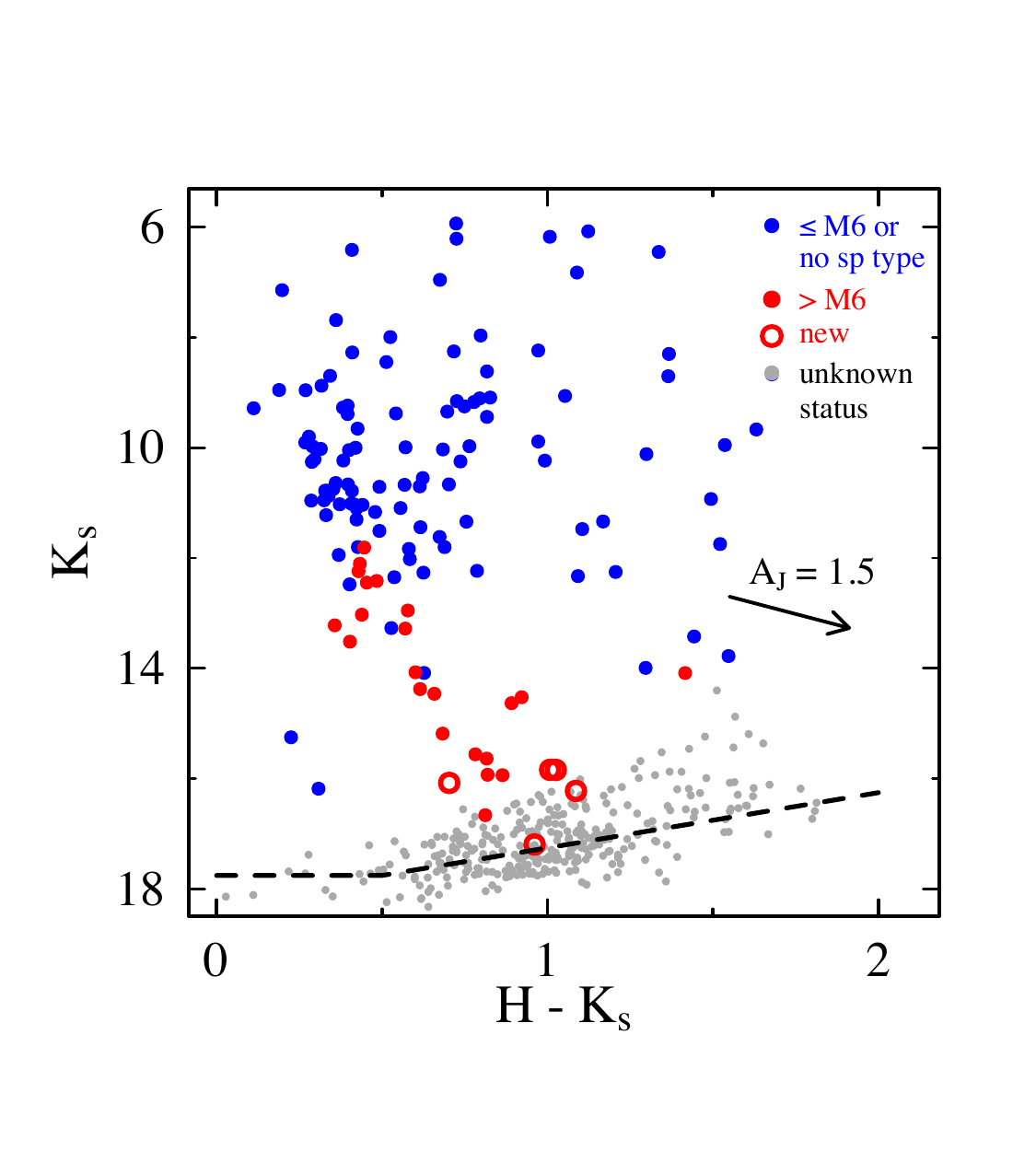}
\caption{
Near-IR color-magnitude diagram of the known members of Cha~I 
within the ISPI fields (large filled and open circles) and the remaining
sources in those fields with unconstrained membership (small gray points).
These data are from ISPI, HAWK-I, and 2MASS. 
The completeness limit of the ISPI images is indicated (dashed line). 
}
\label{fig:near}
\end{figure}

We can characterize the completeness of our new census of members in Cha~I for the 
ISPI fields using a color-magnitude diagram.
In Figure~\ref{fig:near},
we show $K_s$ vs. $H-K_s$ for the 
known members in the ISPI images.
We also include additional sources that are not rejected as field stars 
based on color-magnitude diagrams or proper motions (Section~\ref{sec:selection}).
For extinctions of $A_J < 1.5$, 
there are no remaining sources with undetermined membership in 
these fields down to an extinction-corrected magnitude of $K_s$=15.7.
Using a $K_s$-band bolometric correction for young late-M/early-L dwarfs \citep{fil15},
this limit corresponds to a mass of 6--10~M$_{\rm Jup}$ for ages of 1--3~Myr
according to evolutionary models \citep{bur97,cha00,bar15}.

The new member Cha J11064106$-$7745040 is within the boundaries of the ACS field from \cite{luhm07}
and is within the completeness limits in $A_J$ and $K_s$ that were reported for that field. However, 
it fell in the gap between the ACS detectors, which is why it was not found in that survey. 
\citet{muz15} concluded that their survey for members Cha~I was complete down to a spectral
type of L3. However, we have found two new members within their survey field, 
both of which are earlier than that limit.

\subsection{Distributions of Spectral Type and $M_K$ for an Extinction-limited Sample}
\label{sec:imf}

We have attempted to construct a large sample of members of Cha~I that extends
to low masses and is representative and unbiased in terms of mass by 
considering the known members within the ISPI fields that have $A_J<1.5$.
For most of the members, we have adopted the extinction estimates 
from \cite{luhm07}, \cite{luh08}, and \cite{luhM08}.
Extinctions for members identified since those studies
are measured from our near-IR spectra 
or near-IR colors assuming the intrinsic photospheric values for pre-main sequence stars \citep{luh10}.
Our adopted extinctions are included in Table~\ref{tab:mem}.
The ISPI fields contain 102 members with $A_J<1.5$, 
which corresponds to 42\% of the known members. That sample includes five of
the six new members that we have found. The remaining new member
is outside of the area imaged by ISPI.

\begin{figure}[h]
\centering
\includegraphics[trim = 0mm 120mm 0mm 80mm, clip=true, scale=0.9]{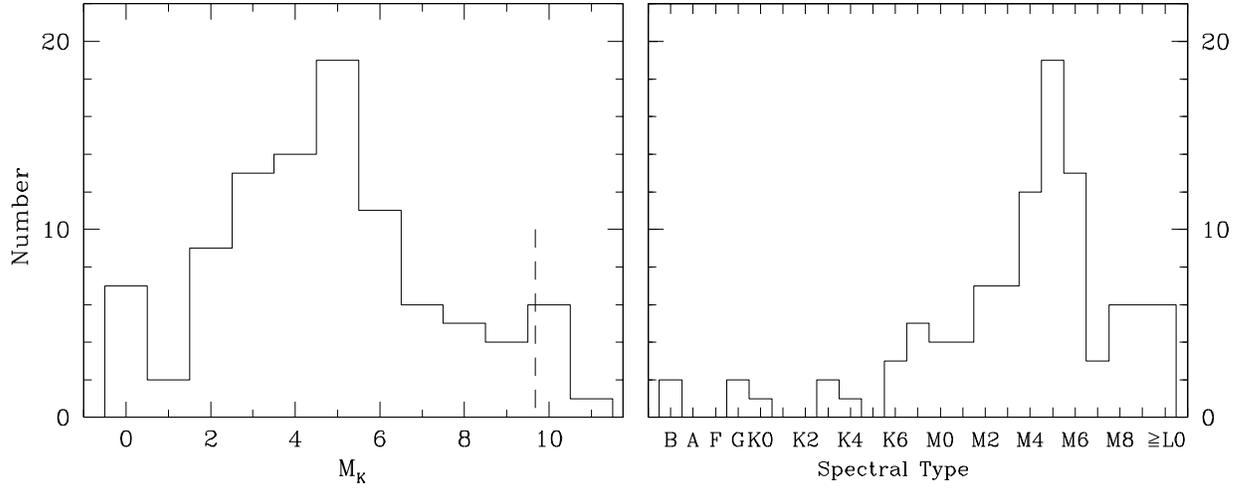}
\caption{
Distribution of extinction-corrected $M_K$ and spectral types for the known 
members of Cha~I that are within the ISPI fields and that have $A_J<1.5$.
This extinction-limited sample is complete down to $M_K=9.7$ (dashed line), 
as indicated by Figure \ref{fig:near}.
}
\label{fig:imf}
\end{figure}

To avoid the uncertainties of mass estimates of young stars that arise from the adopted bolometric corrections,
temperature scales, and evolutionary models,
we characterize the IMF for the extinction-limited sample in the ISPI fields 
using observational proxies for stellar mass, 
namely $M_K$ and spectral type.
In Figure \ref{fig:imf}, 
we show the distributions of spectral types and extinction-corrected $M_K$ for that sample.
Four stars in the sample, 
2MASS J11072022$-$7738111, T33B, T39B, and ESO H$\alpha$ 281, 
are absent from the $M_K$ distribution 
because their $K_s$ photometry is contaminated by an unresolved visual companion. 
2MASS J11095493$-$7635101 and Cha J11081938$-$7731522 are also excluded
from the $M_K$ distribution because they are probably seen in scattered light \citep{luhm07,luh08}.

The distributions of spectral types and $M_K$ in our 
extinction-limited sample in the ISPI fields are similar to
those in the Cha~I fields considered by \cite{luhm07} and in other star-forming clusters \citep{luh16},
exhibiting peaks near M5 and $M_K=5$, respectively. 
Below the peak in $M_K$, the distribution in our new sample is roughly flat down to the completeness
limit of $M_K=9.7$ and extends to $M_K=10.7$ (Cha J11110675$-$7636030), which correspond to masses of
6--10 and 3--6~M$_{\rm Jup}$, respectively, based on evolutionary models for ages of 1--3~Myr
\citep{bur97,cha00,bar15}. 
Thus, the minimum mass of the IMF has not been detected in
these data, and the constraint on that mass is $\lesssim$3--6~M$_{\rm Jup}$.

\section{New Members with Circumstellar Disks}
\label{sec:disk}

\cite{luhM08} and \cite{luh08} measured photometry from mid-IR images from
{\it Spitzer} for the compilation of known members from \cite{luhm08} and used
those data to identify the presence of circumstellar disks. 
We have performed the same exercise for the members found since \cite{luhm08} 
using data from both {\it Spitzer} and the {\it Wide-field Infrared Survey
Explorer} \citep[{\it WISE};][]{wri10}.
Those measurements are compiled in Table \ref{tab:mid}. Because CT Cha B 
is not resolved by {\it Spitzer} or {\it WISE}, we do not include it 
in the table. We measured the {\it Spitzer} data from IRAC images in its
four bands from 3.6--8.0~$\mu$m and from 24~$\mu$m images (denoted as [24])
with the Multiband Imaging Photometer for {\it Spitzer} \citep[][]{rie04}
in the same manner as \cite{luh08} and \cite{luhM08}.
The {\it WISE} photometry was taken from the AllWISE Source Catalog.
Those data were measured in bands at 3.5, 4.5, 12, and 22~$\mu$m,
which are denoted as $W1$, $W2$, $W3$, and $W4$, respectively. 

According to the parameter cc\_flag from the AllWISE
Source Catalog, several of the {\it WISE}
detections in Table \ref{tab:mid} may be contaminated by diffraction spikes or
scattered light halos from nearby bright sources. However, none of those
detections (which are in $W1$ and $W2$) show any such contamination based on
visual inspection of the images.
All of the {\it WISE} sources are consistent with a point source based on the
ext\_flg parameter with the exception of Cha J11105772$-$7714570, which is
blended with another object. Resolved IRAC photometry is available for it.
Although the $W1$ and $W2$ bands are roughly similar to [3.6] and [4.5]
from IRAC, respectively, photometry in those bands can differ noticeably for
cool dwarfs \citep{fil15}, as illustrated by the fact that the $W1$ magnitudes
for the late-M/early-L objects in Table \ref{tab:mid} are systematically larger
than the [3.6] data.

\begin{figure}[h]
\centering
\includegraphics[trim = 0mm 0mm 0mm 0mm, clip=true, scale=0.8]{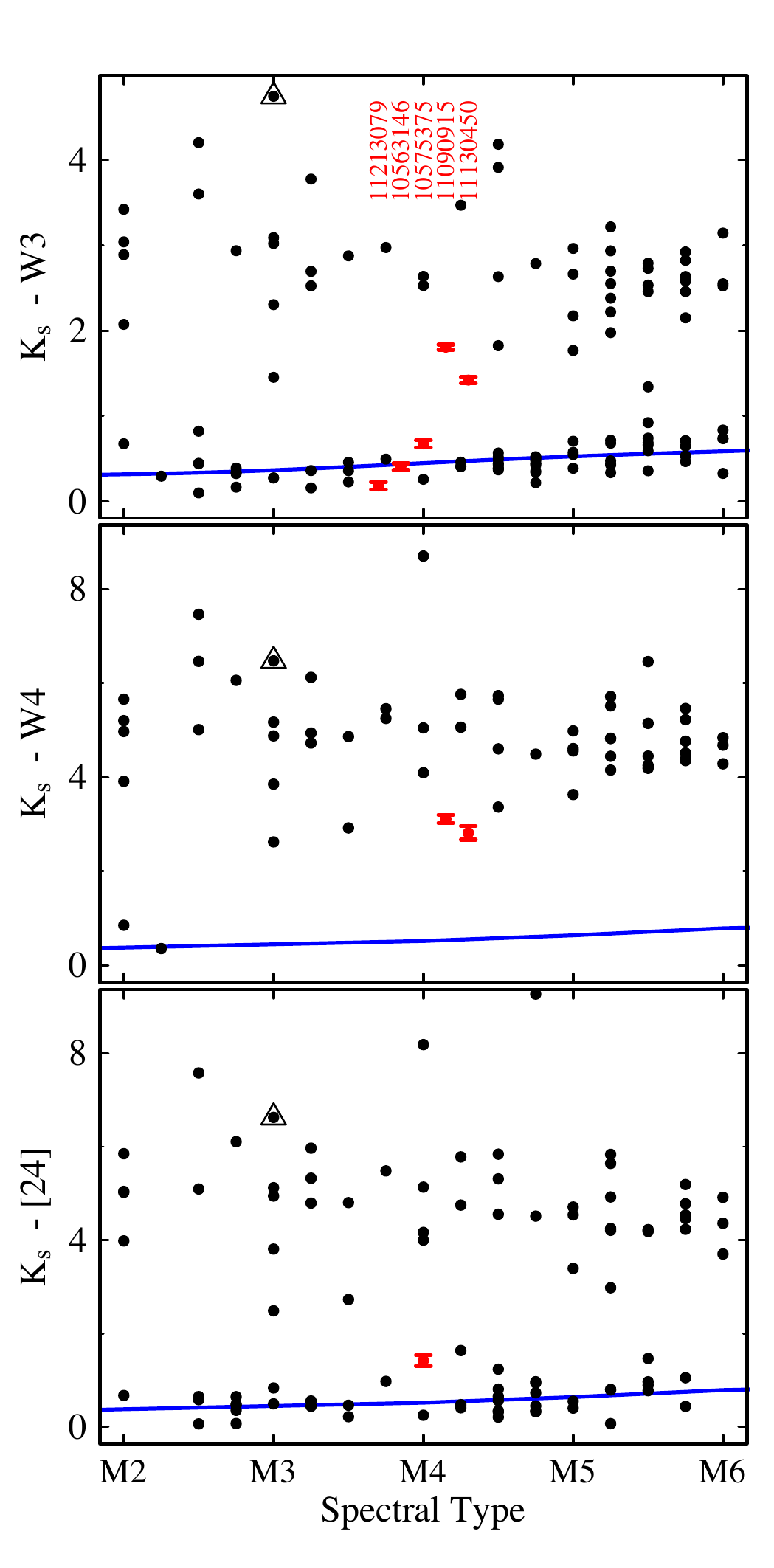}
\caption{
Extinction-corrected mid-IR colors versus spectral type for mid-M members of
Cha~I (filled circles). Most class~I objects (protostars)
are redder than the boundaries of these diagrams; the one that does
appear within their limits is marked (open triangle).
For the members that have been found since
\cite{luhm08} (Table~\ref{tab:mid}), we have included the errors in their
colors and labels with the right ascension component of their names.
Those objects have spectral types of M4, but are plotted at slightly different
spectral types from M3.7--M4.3 so that they can be labeled.
For each color, we have indicated the typical colors of young
stellar photospheres \citep[lines,][Section~\ref{sec:disk}]{luh10}.
Two of the M4 members of Cha~I found since \cite{luhm08} exhibit large
excesses relative to photospheric colors, indicating the presence
of circumstellar disks.
}
\label{fig:disk2}
\end{figure}

To determine whether the members of Cha~I in Table \ref{tab:mid}
have color excesses from disks in the {\it Spitzer} and {\it WISE}
photometry, we begin by plotting diagrams of $K-W3$, $K-W4$, and $K-[24]$
versus spectral type in Figure~\ref{fig:disk2} for a range of spectral types
centered on M4, which is the type of all stars in Table~\ref{tab:mid}
with such data.
In the diagram with $K_s-[24]$, we have included a line that represents
the typical colors of young stellar photospheres from \citet{luh10}.
Because the $W4$ and [24] bands are similar, we show that same line
with the data in $K_s-W4$. For $K_s-W3$, we mark the photospheric sequence
with a fit to the colors of diskless members of Taurus \citep{luh10,esp14}.
Each of the diagrams with $K-W3$ and $K-[24]$ contains a well-defined blue
sequence that closely follows the lines representing typical photospheric
colors. Most M-type members that lack disks were not detected in $W4$, so a
photospheric sequence among Cha~I members is absent from the diagram of
$K_s-W4$. 
2MASS J11130450$-$7534369 and 2MASS J11090915$-$7553477 exhibit significant
color excesses in $K_s-W3$ and $K_s-W4$ relative to photospheric values.
The remaining three M4 members from Table~\ref{tab:mid} do not have excesses.

\begin{figure}[h]
\centering
\includegraphics[trim = 0mm 0mm 0mm 0mm, clip=true, scale=0.8]{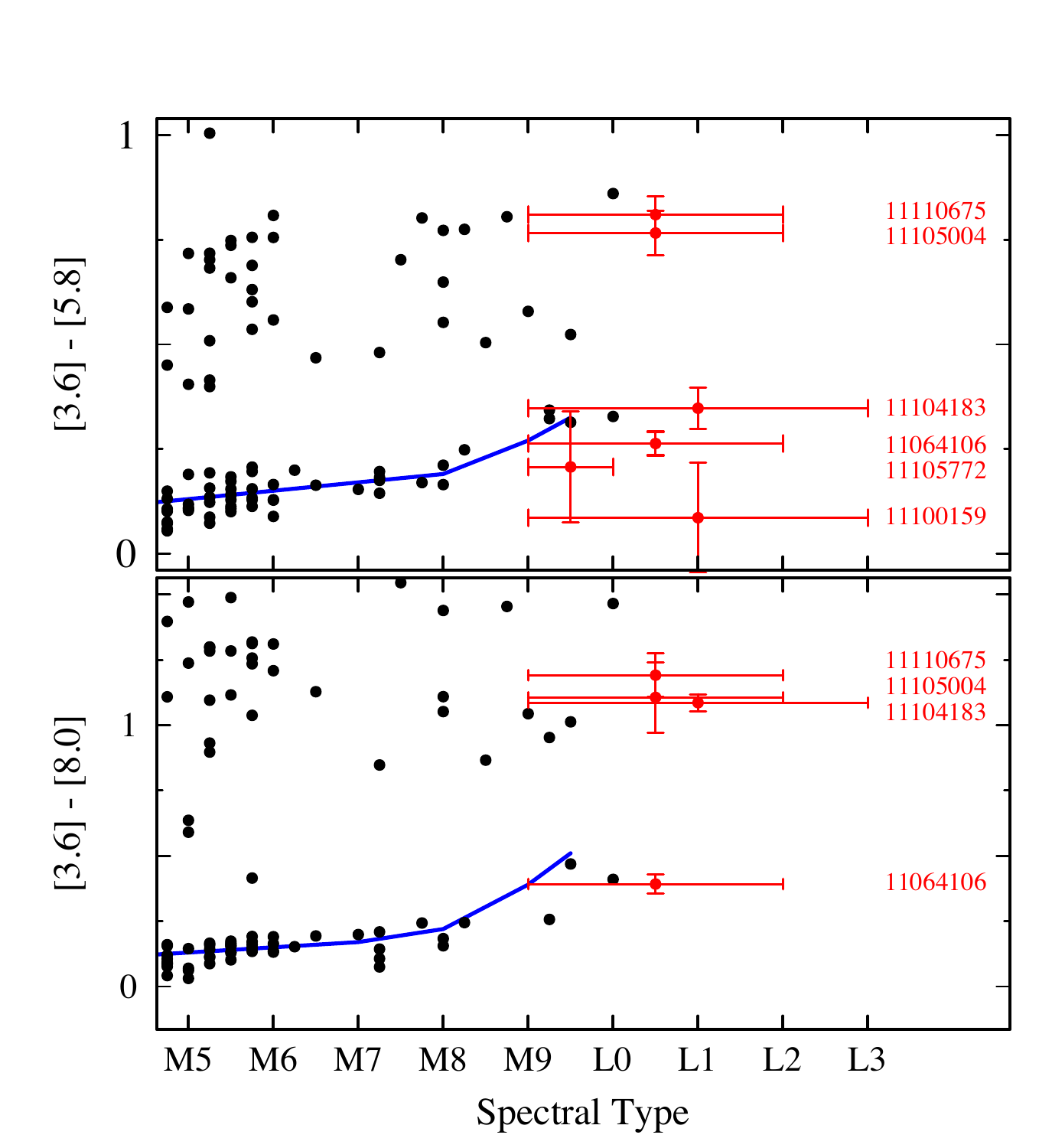}
\caption{
Extinction-corrected mid-IR colors versus spectral type for the coolest known 
members of Cha~I (filled circles).  Some members of Cha~I are too
red to appear within the boundaries of these diagrams, including all of the
known class~I objects. For the members that have been found since
\cite{luhm08} (Table~\ref{tab:mid}), we have plotted the errors in their
spectral types and colors and labels with the right ascension component
of their names.  For each color, we have indicated the typical 
colors for young stellar photospheres at $\leq$M9.5 \citep[lines,][]{luh10}.
Three of the new members of Cha~I are redder than the photospheric sequences
at $\leq$M9.5, which may indicate the presence of disks. However, since the
objects may be later than M9.5 and the photospheric colors are poorly defined at
those types, we cannot determine conclusively whether color excesses from
disks are present.
}
\label{fig:disk}
\end{figure}

To identify excess emission among the coolest new members of Cha~I, we plot
$[3.6]-[5.8]$ and $[3.6]-[8.0]$ as a function of spectral type in
Figure~\ref{fig:disk}. 
We have included lines that represent
the typical colors of young photospheres for spectral types of $\leq$M9.5
\citep{luh10}, which coincide with the sequence of bluest sources in Cha~I. 
Among the six members from Table~\ref{tab:mid} that have IRAC data,
Cha J11110675$-$7636030 (M9--L2), Cha J11105004$-$7721535 (M9--L2), and
Cha J11104183$-$7633064 (M9--L3) have values of $[3.6]-[8.0]$ that are 
significantly redder than the photospheric sequence at $\leq$M9.5.
The same is true for the first two objects in $[3.6]-[5.8]$.
Those red colors may indicate the presence of disks.  However, those three
objects could be as late as L2 or L3 given the uncertainties
in our classifications, and the photospheric values of the IRAC colors
are poorly defined at those types, so we cannot determine definitively whether
they have color excesses from disks. In other words, it is possible that
$[3.6]-[5.8]$ and $[3.6]-[8.0]$ for young photospheres become much redder
at types later than M9.5, in which case the observed colors for
Cha J11110675$-$7636030, Cha J11105004$-$7721535, and Cha J11104183$-$7633064
could be entirely photospheric.
We note that these three objects are not detected in the {\it WISE}
and {\it Spitzer} bands longward of 10~\micron. For instance, the detection
limit of the MIPS data is roughly $[24]\sim10$, which corresponds to limits of
$K_s-[24]<\sim6$--7 for those objects. Thus, the non-detections at longer
wavelengths do not place useful constraints on the presence of disks. 

For all of the objects that exhibit excess emission in Figures~\ref{fig:disk2}
and \ref{fig:disk}, the sizes of the excesses are similar to those
of members of Cha~I \citep{luh08,luhm08} and Taurus \citep{esp14} that have
class~II spectral energy distributions \citep[star+disk,][]{lw84,lad87}.

Through our analysis of the mid-IR photometry in Table \ref{tab:mid}, we have
found that three of our new members may have excess emission that indicates
the presence of disks. If excesses are present, they would provide additional
evidence of the youth and membership of those objects.
Two of these members, Cha J11105004$-$7721535 and Cha J11110675$-$7636030,
would be the faintest known members in extinction-corrected $M_K$ that show
evidence of disks \citep{luh08,luhM08}. In addition, the latter could be the
faintest known brown dwarf in nearby star-forming regions with a 
detection of a disk \citep{luhM12,esp14,luh16}. Thus, it may be the
least-massive known brown dwarf with a circumstellar disk (3--6~M$_{\rm Jup}$).
However, additional data (e.g., photometry at longer wavelengths) are needed 
to determine conclusively whether these objects harbor disks.

\section{Conclusion}

We have conducted a survey for planetary-mass brown dwarfs in the Cha~I
star-forming region using proper motions and photometry measured from optical
and near-IR imaging with {\it HST}, {\it Spitzer}, and ground-based telescopes. 
Through spectroscopy of the resulting candidate members,
we have discovered six new late-type members, which include three of the five
faintest known members in extinction-corrected $M_K$.
One of these objects, Cha J11110675$-$7636030, is the faintest known member, 
which has a mass of 3--6 M$_{\rm Jup}$ according to evolutionary models. 
Three of the new members have red mid-IR colors that may indicate the presence
of disks, but these detections of disks are not definitive given the
uncertainties in the spectral types of the objects and the uncertainties in
the photospheric colors of young brown dwarfs later than L0. If its red color
is due to a disk, Cha J11110675$-$7636030 would be a contender for the
least-massive known brown dwarf with a disk. Given that we have found objects
below the completeness limit for our survey (6--10~M$_{\rm Jup}$), deeper
imaging is required to measure the minimum mass of the IMF in Cha~I.

\acknowledgements
We thank K. Allers for providing the modified version of Spextool for use with
ARCoIRIS data. We acknowledge support from grants AST-1208239 and AST-1313029 from the
National Science Foundation (NSF)
and grant GO-11695 from the Space Telescope Science Institute.
EEM acknowledges support from the NASA NExSS program. 
The {\it Spitzer Space Telescope} and the IPAC Infrared Science Archive (IRSA)
are operated by JPL and Caltech under contract with NASA.
The NASA/ESA {\it Hubble Space Telescope} is operated by the 
Space Telescope Science Institute and the Association of Universities
for Research in Astronomy (AURA), Inc., under NASA contract NAS 5-26555.
2MASS is a joint project of the University of
Massachusetts and the Infrared Processing and Analysis Center (IPAC) at
Caltech, funded by NASA and the NSF.
The Center for Exoplanets and Habitable Worlds is supported by the
Pennsylvania State University, the Eberly College of Science, and the
Pennsylvania Space Grant Consortium. 
We thank the CTIO staff for help with the DECam observing run, especially Claudio Aguilera, Alberto Alvarez, Kadur Flores, David James, Leonor Opazo, Javier Rojas, and Hernan Tirado. 
The data at CTIO were obtained 
through program 2015A-0175 for National Optical Astronomy Observatory (NOAO).
CTIO and NOAO are operated by the AURA under a cooperative agreement with the
NSF. 
The data at Gemini were obtained through program GS-2015A-Q-64.
The Gemini Observatory is operated by AURA, under a cooperative agreement with the NSF on behalf of the Gemini partnership: the National Science Foundation (United States), the National Research Council (Canada), CONICYT (Chile), Ministerio de Ciencia, Tecnolog\'{i}a e Innovaci\'{o}n Productiva (Argentina), and Minist\'{e}rio da Ci\^{e}ncia, Tecnologia e Inova\c{c}\~{a}o (Brazil).
This project used data obtained with the Dark Energy Camera (DECam), which was constructed by the Dark Energy Survey (DES) collaboration.
Funding for the DES Projects has been provided by 
the U.S. Department of Energy, the U.S. National Science Foundation, 
the Ministry of Science and Education of Spain, 
the Science and Technology Facilities Council of the United Kingdom, 
the Higher Education Funding Council for England, 
the National Center for Supercomputing Applications at the University of Illinois at Urbana-Champaign, 
the Kavli Institute of Cosmological Physics at the University of Chicago, 
the Center for Cosmology and Astro-Particle Physics at the Ohio State University, 
the Mitchell Institute for Fundamental Physics and Astronomy at Texas A\&M University, 
Financiadora de Estudos e Projetos, Funda{\c c}{\~a}o Carlos Chagas Filho de Amparo {\`a} Pesquisa do Estado do Rio de Janeiro, 
Conselho Nacional de Desenvolvimento Cient{\'i}fico e Tecnol{\'o}gico and the Minist{\'e}rio da Ci{\^e}ncia, Tecnologia e Inovac{\~a}o, 
the Deutsche Forschungsgemeinschaft, 
and the Collaborating Institutions in the Dark Energy Survey. 
The Collaborating Institutions are 
Argonne National Laboratory, 
the University of California at Santa Cruz, 
the University of Cambridge, 
Centro de Investigaciones En{\'e}rgeticas, Medioambientales y Tecnol{\'o}gicas-Madrid, 
the University of Chicago, 
University College London, 
the DES-Brazil Consortium, 
the University of Edinburgh, 
the Eidgen{\"o}ssische Technische Hoch\-schule (ETH) Z{\"u}rich, 
Fermi National Accelerator Laboratory, 
the University of Illinois at Urbana-Champaign, 
the Institut de Ci{\`e}ncies de l'Espai (IEEC/CSIC), 
the Institut de F{\'i}sica d'Altes Energies, 
Lawrence Berkeley National Laboratory, 
the Ludwig-Maximilians Universit{\"a}t M{\"u}nchen and the associated Excellence Cluster Universe, 
the University of Michigan, 
{the} National Optical Astronomy Observatory, 
the University of Nottingham, 
the Ohio State University, 
the University of Pennsylvania, 
the University of Portsmouth, 
SLAC National Accelerator Laboratory, 
Stanford University, 
the University of Sussex, 
and Texas A\&M University.
This paper has been approved for Unlimited External Release (URS266045).

{\it Facilities: } \facility{Blanco (DECam, ARCoIRIS)}, 
\facility{SOAR (Goodman)},
\facility{Gemini:South (Flamingos-2)},
\facility{HST (ACS, WFP2)},
\facility{Spitzer (IRAC)},
\facility{VLT (HAWK-I)}

\clearpage

\begin{deluxetable}{ll}
\tabletypesize{\scriptsize}
\tablewidth{0pt}
\tablecaption{Members of Chamaeleon I \label{tab:mem}}
\tablehead{
\colhead{Column Label} &
\colhead{Description}}
\startdata
Name & Source name\tablenotemark{a} \\
OtherNames & Other source name \\
SpType & Adopted spectral type \\
r\_SpType & Spectral type reference\tablenotemark{b} \\
pmRA & IRAC relative proper motion in right ascension \\
e\_pmRA & Error in pmRA \\
pmDec & IRAC relative proper motion in declination \\
e\_pmDec & Error in pmDec \\
flag & Flag on IRAC relative proper motions\tablenotemark{c} \\
Aj & Extinction in $J$ \\
r\_Aj & Extinction in $J$ reference\tablenotemark{d} \\
Jmag & $J$ magnitude \\
e\_Jmag & Error in Jmag \\
Hmag & $H$ magnitude \\
e\_Hmag & Error in Hmag \\
Ksmag & $K_s$ magnitude \\
e\_Ksmag & Error in Ksmag \\
JHKref & JHK reference\tablenotemark{e} 
\enddata
\tablecomments{This table is available in its entirety in a machine-readable form in the online
journal.}
\tablenotetext{a}{Coordinate-based identifications from the 2MASS Point Source Catalog 
when available. Otherwise, identifications are based on the 
coordinates measured in this work or \cite{luhm08}.}
\tablenotetext{b}{1 = \cite{luhm08}, references therein; 2 = \cite{fra15}; 3 = this work;
4 = revised classification of spectrum from \cite{luh06}; 5 = revised classification of 
spectrum from \cite{luhm07}.}
\tablenotetext{c}{
nodet = non-detection;
sat = saturated;
out = outside of overlapping multi-epoch IRAC images;
unres = too close to a brighter star to be detected;
err = motion error above our adopted thresholds. 
}
\tablenotetext{d}{
1 = \cite{luhm07};
2 = \cite{luhM08};
3 = \cite{luh08};
sp = derived from spectrum with spectral templates from \cite{luh16b};
J-H = derived from the $J-H$ color assuming photospheric near-infrared colors \citep{luh10}.
}
\tablenotetext{e}{
2 = 2MASS Point Source Catalog; 
l = \cite{luh05};
i = ISPI data from this work;
h = HAWK-I data from this work;
c = \cite{car02}.
}
\end{deluxetable}

\begin{deluxetable}{ccll}
\tabletypesize{\scriptsize}
\tablecaption{IRAC Observations of Chamaeleon I \label{tab:epochs}}
\tablehead{
\colhead{AOR} & \colhead{PID} & \colhead{P. I.} & \colhead{Epoch}
}
\startdata
3651328 & 6 & G. Fazio & 2004.5\\
3955968 & 36 & G. Fazio & 2004.7\\
3960320 & 37 & G. Fazio & 2004.5\\
5105920 & 139 & N. Evans & 2004.5\\
5662720 & 173 & N. Evans & 2004.3\\
5662976 & 173 & N. Evans & 2004.5\\
6526208 & 36 & G. Fazio & 2004.1\\
12620032 & 36 & G. Fazio & 2005.6\\
12620544 & 36 & G. Fazio & 2005.6\\
18366720 & 30540 & J. Houck & 2006.5\\
18366976 & 30540 & J. Houck & 2006.5\\
18367232 & 30540 & J. Houck & 2006.5\\
18367744 & 30540 & J. Houck & 2006.5\\
18368000 & 30540 & J. Houck & 2006.5\\
18368256 & 30540 & J. Houck & 2006.5\\
18368512 & 30540 & J. Houck & 2006.5\\
18368768 & 30540 & J. Houck & 2006.5\\
18369024 & 30540 & J. Houck & 2006.5\\
18369280 & 30540 & J. Houck & 2006.5\\
18369536 & 30540 & J. Houck & 2006.5\\
18369792 & 30540 & J. Houck & 2006.5\\
18370048 & 30540 & J. Houck & 2006.5\\
18370304 & 30540 & J. Houck & 2006.5\\
18374400 & 30540 & J. Houck & 2006.5\\
19986432 & 30574 & L. Allen & 2007.4\\
19992832 & 30574 & L. Allen & 2007.4 \\
20006400 & 30574 & L. Allen & 2007.4\\
20012800 & 30574 & L. Allen & 2007.4\\
20014592 & 30574 & L. Allen & 2007.4 \\
20015104 & 30574 & L. Allen & 2007.4\\
47089152 & 90071 & A. Kraus & 2013.6 \\
47089664 & 90071 & A. Kraus & 2013.6\\
47090176 & 90071 & A. Kraus & 2013.6\\
47090688 & 90071 & A. Kraus & 2013.6\\
47091200 & 90071 & A. Kraus & 2013.6\\
47091712 & 90071 & A. Kraus & 2013.6 \\
47092224 & 90071 & A. Kraus & 2013.6\\
47092736 & 90071 & A. Kraus & 2013.6 \\
47093248 & 90071 & A. Kraus & 2013.6
\enddata
\end{deluxetable}

\begin{deluxetable}{lcc}
\tablecolumns{3}
\tabletypesize{\scriptsize}
\tablewidth{0pt}
\tablecaption{New Members of Chamaeleon~I\label{tab:spec}}
\tablehead{
\colhead{Source Name\tablenotemark{a}} &
\colhead{Spectral} &
\colhead{Spectrograph} \\ 
& \colhead{Type} 
}
\startdata
Cha J11064106$-$7745040 & M9--L2 & FLAMINGOS-2 \\ 
Cha J11104183$-$7633064 & M9--L3 & FLAMINGOS-2 \\ 
Cha J11105004$-$7721535 & M9--L2 & FLAMINGOS-2 \\ 
Cha J11105772$-$7714570 & M9.5 & FLAMINGOS-2 \\ 
Cha J11110675$-$7636030 & M9--L2 & FLAMINGOS-2 \\ 
2MASS J11124771$-$7737547 & M8 & FLAMINGOS-2 
\enddata
\tablenotetext{a}{Coordinate-based identifications from the 2MASS Point Source
Catalog when available. Otherwise, identifications are based on the coordinates 
measured in this work.}
\end{deluxetable}

\clearpage
\begin{turnpage}
\begin{deluxetable}{lccccccccccc}
\tablecolumns{11}
\tabletypesize{\scriptsize}
\tablewidth{0pt}
\tablecaption{Mid-IR Photometry for Members of Chamaeleon~I Found
Since \cite{luhm08}\label{tab:mid}}
\tablehead{
\colhead{Source Name} &
\colhead{$W1$} &
\colhead{$W2$} &
\colhead{$W3$} &
\colhead{$W4$} &
\colhead{[3.6]} &
\colhead{[4.5]} &
\colhead{[5.8]} &
\colhead{[8.0]} &\colhead{[24]} & \colhead{Excess?} \\ 
& \colhead{(mag)} & \colhead{(mag)} & \colhead{(mag)} & \colhead{(mag)} & \colhead{(mag)} & \colhead{(mag)} & \colhead{(mag)} & \colhead{(mag)} & \colhead{(mag)} 
}
\startdata
2MASS J10563146$-$7618334 & $10.07\pm0.02$ & $9.86\pm0.02$ & $9.73\pm0.03$ & $9.73\pm0.03$\tablenotemark{a} &
	out & out & out & out & out & N \\
2MASS J10575375$-$7724495 & $10.28\pm0.12$ & $10.19\pm0.02$ & $9.77\pm0.04$ & $8.54\pm0.25$\tablenotemark{a} &
	$10.23\pm0.02$ & $10.18\pm0.02$ & out & out & $9.00\pm0.11$ & N \\
Cha J11064106$-$7745040 & $15.31\pm0.03$ & $14.85\pm0.05$ & $>$12.4 & $>$9.4 &
	$14.86\pm0.02$ & $14.71\pm0.02$ & $14.58\pm0.02$ & $14.46\pm0.03$ & \nodata & N \\
2MASS J11090915$-$7553477 & $10.29\pm0.02$ & $10.05\pm0.02$ & $8.66\pm0.02$ & $7.33\pm0.08$ &
	out & out & out & out & out & Y \\
Cha J11100159$-$7738052\tablenotemark{b} & $16.57\pm0.08$ & $16.25\pm0.18$ & $>$13.2 & $9.06\pm0.33$\tablenotemark{a} &
	$15.88\pm0.02$ & $15.67\pm0.02$ & $15.79\pm0.13$ & \nodata & \nodata & N \\
Cha J11104183$-$7633064 & $15.15\pm0.03$ & $14.59\pm0.05$ & $>$12.7 & $>$8.9 &
	$14.72\pm0.02$ & $14.38\pm0.02$ & $14.36\pm0.04$ & $13.62\pm0.02$ & \nodata & Y? \\
Cha J11105004$-$7721535 & $15.66\pm0.11$ & $15.09\pm0.05$ & $>$12.8 & $>$9.4 &
	$15.17\pm0.02$ & $14.81\pm0.02$ & $14.40\pm0.05$ & $14.06\pm0.13$ & \nodata & Y? \\
Cha J11105772$-$7714570& $15.18\pm0.40$\tablenotemark{c} & $14.75\pm0.04$\tablenotemark{c} & $>$12.5 & $>$9.3 &
	$15.14\pm0.02$ & $14.92\pm0.02$ & $14.93\pm0.13$ & \nodata & \nodata & N \\
Cha J11110675$-$7636030 & \nodata & \nodata & \nodata & \nodata &
	$15.94\pm0.02$ & $15.50\pm0.02$ & $15.12\pm0.04$ & $14.74\pm0.08$ & \nodata & Y? \\
2MASS J11124771$-$7737547 & $13.70\pm0.03$ & $13.33\pm0.03$ & $12.16\pm0.24$\tablenotemark{d} & $>$9.5 &
	out & out & out & out & \nodata & N \\
2MASS J11130450$-$7534369 & $10.59\pm0.02$ & $10.30\pm0.02$ & $9.33\pm0.03$ & $7.93\pm0.14$ &
	out & out & out & out & out & Y \\
2MASS J11213079$-$7633351 & $9.78\pm0.02$ & $9.69\pm0.02$ & $9.62\pm0.04$ & $>$9.1 &
	out & out & out & out & out& N 
\enddata
\tablecomments{ 
Ellipses and ``out" indicate measurements that are absent because of non-detection
and a position outside of the camera's the field of view, respectively. 
}
\tablenotetext{a}{Detection is false or unreliable based on visual inspection.}
\tablenotetext{b}{Photometry for Cha J11100159$-$7738052 is not reported in
the AllWISE Source Catalog, so we list the data from the {\it WISE}
All-Sky Source catalog.}
\tablenotetext{c}{Blended with another star.}
\tablenotetext{d}{The AllWISE Source Catalog reports a detection but it
appears to be offset from the detections in $W1$ and $W2$.}
\end{deluxetable}
\end{turnpage}
\clearpage
\global\pdfpageattr\expandafter{\the\pdfpageattr/Rotate 90}

\global\pdfpageattr\expandafter{\the\pdfpageattr/Rotate 360}

\end{document}